\documentclass[]{interact}

\usepackage[outdir=./]{epstopdf}

\usepackage[numbers,sort&compress,merge]{natbib}

\bibpunct[, ]{[}{]}{,}{n}{,}{,}
\renewcommand\bibfont{\fontsize{10}{12}\selectfont}

\theoremstyle{plain}

\theoremstyle{definition}

\theoremstyle{remark}

\usepackage[version=3]{mhchem} 
\usepackage[hang]{subfigure}
\usepackage{xcolor}
\usepackage{dirtytalk}
\usepackage{verbatim}
\definecolor{darkblue}{HTML}{003D6D}
\usepackage{tikz}
\usepackage{graphicx}
\usepackage{float}
\usepackage{blindtext}
\usepackage[export]{adjustbox}
\usepackage{changepage}
\usepackage{hyperref}
\usepackage{colortbl}

\newlength{\offsetpage}
{\end{adjustwidth}}

\usetikzlibrary{positioning}

\begin{document}


\title{HOAX: A Hyperparameter Optimization Algorithm Explorer for Neural Networks}

\author{
\name{Albert Thie\textsuperscript{a}, Maximilian F.S.J. Menger\textsuperscript{a} and Shirin Faraji*\textsuperscript{a}\thanks{CONTACT Shirin Faraji  Email: s.s.faraji.rug.nl}}
\affil{\textsuperscript{a}Zernike Institute for Advanced Materials, Faculty of Science and Engineering, University of Groningen, Nijenborgh 4, 9747AG Groningen The Netherlands. }
}

\maketitle


\begin{keywords}
Quantum Chemistry, Machine Learning, Neural Networks, Hyperparameter Optimization
\end{keywords}



\begin{abstract}
Computational chemistry has become an important tool to predict and understand molecular properties and reactions. Even though recent years have seen a significant growth in new algorithms and computational methods that speed up quantum chemical calculations, the bottleneck for trajectory-based methods to study photo-induced processes is still the huge number of electronic structure calculations. In this work, we present an innovative solution, in which the amount of electronic structure calculations is drastically reduced, by employing machine learning algorithms and methods borrowed from the realm of artificial intelligence. However, applying these algorithms effectively requires finding optimal hyperparameters, which remains a challenge itself. Here we present an automated user-friendly framework, HOAX, to perform the hyperparameter optimization for neural networks, which bypasses the need for a lengthy manual process. The neural network generated potential energy surfaces (PESs) reduces the computational costs compared to the ab initio-based PESs. We perform a comparative investigation on the performance of different hyperparameter optimiziation algorithms, namely grid search, simulated annealing, genetic algorithm, and bayesian optimizer in finding the optimal hyperparameters necessary for constructing the well-performing neural network in order to fit the PESs of small organic molecules. Our results show that this automated toolkit not only facilitates a straightforward way to perform the hyperparameter optimization but also the resulting neural networks-based generated PESs are in reasonable agreement with the ab initio-based PESs.
\end{abstract}


\section{Introduction}
\begin{figure*}
\makebox[\linewidth][c]{
\centering
\includegraphics[width=1\textwidth]{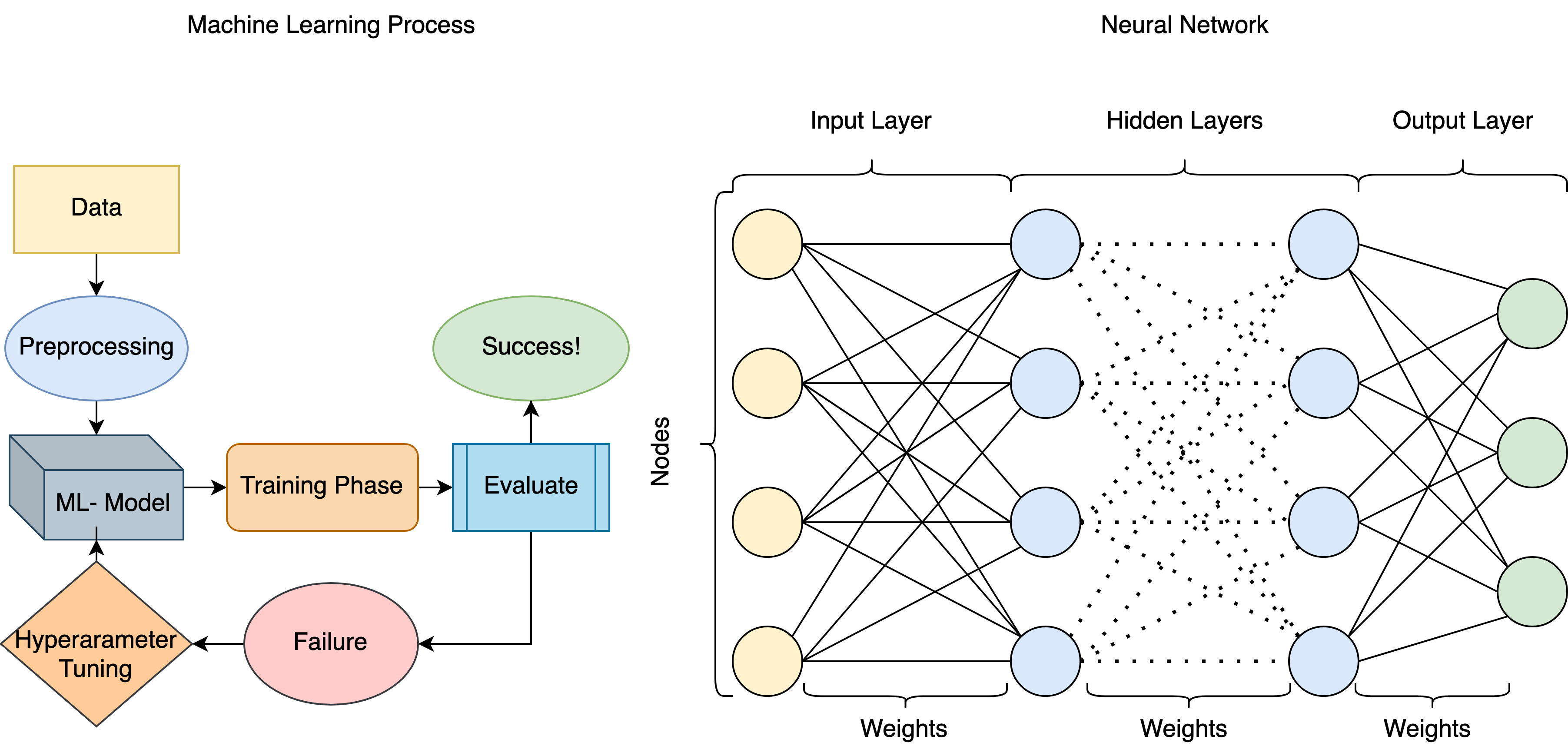}}
\caption{Systematic overview of the process of training a ML model (left) and a neural network architecture (right).}
\label{fig:schemaoverview}
\end{figure*}

The application of machine learning (ML) algorithms has grown to encompass many new fields, such as data analysis of voting records\cite{baldwin2017myths}, enhancing the motor capabilities in robotics\cite{gardner2008structure} and even in creating fictional stories\cite{gervas2006narrative}. In the field of electronic structure theory, ML algorithms have also seen an increased range of applications. ML is applied to approximate density functionals\cite{snyder2012finding}, generating the ground-\cite{schnet,PhysRevLett.98.146401} and excited-state potential energy surfaces (PESs)\cite{westermayr2020machine,surfacehoppingkernel}. A major challenge in applying any ML model is choosing a suitable ML algorithm and the corresponding optimal hyperparameters\cite{moderndeeplearning,machinelearningoverview} (see Figure \ref{fig:schemaoverview}), for which various optimization algorithms exist\cite{bergstra2011algorithms}.\\

Currently, neural networks are among the most commonly applied ML algorithms in computational chemistry\cite{machinelearningoverview}. Most of the applications are centered around describing the ground-state molecular properties. For example, SchNet uses a deep learning architecture that is applied to the QM9 database\cite{schnet}, which consists of 134k stable small organic molecules made of \ce{C}, \ce{H}, \ce{O}, \ce{N}, and \ce{F} atoms\cite{ramakrishnan2014quantum}, to predict a wide range of molecular properties across chemical space. PauliNet uses deep neural network quantum quantum Monte Carlo based approach that is capable to achieve nearly exact solutions of the electronic Schr{\"o}dinger equation\cite{Hermann2019DeepNN}. Other examples include systems that use locality to restrict the input space for the neural network\cite{locality}, a hierarchically interacting particle neural network which learns to transform the input space into terms describing their interactions\cite{doi:10.1063/1.5011181} or networks explicitly designed to be covariant\cite{Anderson2019CormorantCM}.
Message passing networks are another successful approach that uses graph neural networks to make accurate predictions of the quantum mechanical properties of small organic molecules\cite{duvenaud2015convolutional,graph1,Schtt2017}. Other studies have used kernel-based ML methods, such as kernel-based regression methods to predict various molecular properties, such as PES(s), normal modes, internal energy, and heat capacity\cite{doi:10.1063/1.5053562,doi:10.1063/1.5020710,kernelLilienfeld2015,rupp2015machinekernel1}.\\

For excited-state dynamics, the SchNarc\cite{schnarc} approach combines SchNet\cite{schnet} and the non-adiabatic dynamics package SHARC\cite{sharcmd21}. This results in a deep learning approach that can run excited-state dynamics simulations based on the learned important properties, such as non-adiabatic couplings, gradients, Hessians, and spin–orbit couplings, to further simplify such simulations.  
Neural networks have also been used to generate surfaces for the treatment of the excited states of Formaldehyde\cite{KochMCTDH}, long-term simulation of excited-states dynamics replacing force field methods\cite{longtermforcefield}, and using kernel ridge regression to provide properties for decoherence-corrected fewest switches surface hopping\cite{surfacehoppingkernel}. A good overview of the recent works in this field can be found in Ref \cite{westermayr2019machine}.\\

Although these groundbreaking works have illustrated the potential of ML algorithms to accelerate and improve molecular simulations, a major challenge from the user perspective, besides accurate and comprehensive training data and the corresponding computational costs, is finding the suitable ML algorithm and the associated hyperparameters out of the wide range of existing ML methods.
Kernel-based ML methods have the advantage that the hyperparameters can be determined a priori, using the descriptors without any reference to the target property\cite{scholkopfkernel}. Although this remains a challenge for neural network modeling, it has been shown in a comparative research that neural networks can produce better results, given the optimal hyperparameters are found\cite{algcomparison}. The open-source PES-Learn package uses the hyperparameter optimizer (HyperOpt) to find the hyperparameters for neural networks and Gaussian processes\cite{peslearn,bergstra2015hyperopt}. To the best of our knowledge, comparing hyperparameter optimization methods to find optimal hyperparameters to construct a reliable neural network has not been investigated, which motivates the idea of introducing an automated procedure for hyperparameter optimization that is the main focus of the present work.\\


In this work, we investigate the performance of neural networks in predicting the PESs of a set of small organic molecules, namely, \ce{SO_2}, Pyrazine, Pyrrole, and Furan. We investigate the performance of various derivative-free hyperparameter optimization schemes\cite{rios2013derivative}, as we cannot directly compute a gradient for the hyperparameter optimization. We therefore leave out gradient dependent optimization algorithms such as gradient descent\cite{gradientdescent}. We perform a cross-comparison of different optimization methods, namely grid search (GS)\cite{gridsearchcv}, simulated annealing (SA)\cite{simulatedanealing} genetic algorithms (GA)\cite{hooft2021discovering,dinggenetic} and bayesian optimization (BO) \cite{bayesmachine}, to explore the relationship between the performance of neural networks and the used optimization technique. We also investigate the relationship between the molecule that is modelled and the corresponding hyperparameters, as well as the size of the training data.\\

This paper is organized as follows; we first discuss the structure of the neural networks and the hyperparameters that are employed. Secondly, we describe the hyperparameter optimization algorithms that are used to find the optimal neural network models. Thirdly, we compare the results of hyperparameter optimizers and their respective neural network models in constructing the PESs of \ce{SO2}, Pyrazine, Pyrrole, and Furan. Finally, we discuss the effectiveness and possible extension of the package.
\section{Theory}

\subsection{Neural networks}

Neural networks are comprised of layers of independent discriminators, named nodes\cite{gurney2014introduction}, containing an input layer, one or more hidden layers, and an output layer. Each node connects to other nodes and has an associated weight (see Figure \ref{fig:schemaoverview}). 
These weights are updated by training the neural network. The performance of a neural network is determined by the hyperparameters of the network, e.g., the number of nodes per layer, the number of hidden layers, the learning rate to only name a few, which are typically preselected by the user. In this work, we present the hyperparameter optimization using a set of fixed and flexible hyperparameters. The fixed hyperparameters can be set by the user in the configuration file and are unchanged throughout the hyperparameter optimization process. For the flexible hyperparameters, the user defines a range in which the hyperparameter optimization algorithm looks for the optimal hyperparameters. The fixed and flexible hyperparameters that are used in this work can be found in Table \ref{tab1:NNSettings} and are further described in the following.

\begin{table}[H]
\small
    \caption{Overview of the fixed and flexible hyperparameters used in training the neural networks.}
    \label{tab1:NNSettings}
\begin{tabular}{|l|l|}
\hline
\textbf{Fixed}     & \textbf{Flexible} \\ \hline
Activation Function  & Number of Layers \\ \hline
Loss Function & Number of Nodes  \\ \hline
Internal Optimizer & Learning Rate     \\ \hline
Epoch Number   & Batch Size        \\ \hline
\end{tabular}
\end{table}

An important hyperparameter is the activation function, which is applied to the activation $a$ of every node. One can write the activation $a$ of a node as the sum of the inputs $x_i$ for the node with their respective weights $w_i$, together with the bias $\beta$ of the node, as seen in equation \ref{eq:1}

\begin{equation}\label{eq:1}
    a = \sum w_i x_i +\beta
\end{equation}

The most commonly used activation functions are currently supported by the package (see Table \ref{tab2:NNfunctionals}). 
\begin{table}[H]
\small
    \caption{Overview of the different activation functions available in HOAX. Activation functions are applied to the output of each node in the neural network.}
    \label{tab2:NNfunctionals}
\begin{tabular}{|l|l|}
\hline
\textbf{Name}     & \textbf{Formula} \\ \hline
Sigmoid & $\sigma(x) = \frac{1}{1 + e^{-x}}$ \\ \hline
ReLu          & $ReLu(x) = max(0,x)$  \\ \hline
TanH     & $TanH(x) = \frac{e^x - e^{-x}}{e^x + e^{-x}}$  \\ \hline
\end{tabular}
\end{table}

Another important hyperparameter is the loss function, which determines how the error of a neural network is calculated. The error is the difference between the output of the neural network and the training (reference) output, for the given training input. In this research, the error is the difference between the energy in the database and the output of the network for each set of coordinates. Implemented in the package are the mean absolute error (MAE), $\frac{1}{N}\sum_i |E_i|$, the mean squared error (MSE), $\frac{1}{N}\sum_i E^2_i$, and the root mean squared error (RMSE), $ \sqrt{\frac{1}{N}\sum_i E^2_i}$ as different loss functions. Previous works on predicting PESs using neural networks used the MAE or MSE as both the measure when reporting the results on the reference data and in training the neural network\cite{lubbers2018hierarchical,schutt2018schnet,westermayr2020machine}. In this work, we use the MSE for training the neural network, while using the RMSE and MAE for the validation step.\\

The neural network internal optimizer is the algorithm that modifies the neural network weights after receiving the error from the loss function. The HOAX package currently supports two neural network optimizers; i) the stochastic gradient descent (SGD) optimizer and ii) the adaptive moment estimation (ADAM) optimizer\cite{kingma2017adam}.
The SGD uses the gradient of the error multiplied by the learning rate parameter to calculate the changes for the weights connected to the output layer of the network. It then iteratively adjusts the weights in each previous layer, calculating the error of each individual node using backpropagation\cite{gurney2014introduction}. The ADAM optimizer\cite{kingma2017adam} on the other hand changes the weights in the network using an adaptive learning rate, depending on the amount of previous changes for individual weights. It also looks at the average of the gradients of a weight over time and uses this average to change the weights. It can therefore deal with more noisy and sparse data and provides better results and training speed than the SGD optimizer\cite{kingma2017adam}.\\

The epoch number is the maximum number of training cycles performed. An upper bound is required to constrain the training time when the network has not converged to a sufficient minimal error within the predefined training cycles.\\

The flexible parameters in our package are the number of layers, the number of nodes within each layer, the learning rate, and the batch size in the neural network. The learning rate controls the pace at which the optimizer updates the weights with respect to the loss function. It defines how quickly the neural network updates the weights.
The batch size parameter refers to the amount of training examples that are used per training cycle. A batch number of 1 indicates the weights of the model which are updated after each training example. A higher number indicates the amount of training examples for generating an average error, which is used to update the weights.

\subsection{Hyperparameter optimization}

In order to approach the problem of using the right hyperparameters, we can define the problem as optimizing an objective function\cite{Geoffrion1977}. An objective function assumes that we can approach some function $f(x)$ by some approximate function $\widetilde{f}(x)$. We can determine the $\widetilde{f}(x)$ by looking at the difference between the output of the  $\widetilde{f}(x)$ and $f(x)$, for some $x \in X$. Here $X$ is the range that is applied to the approximate function. We call the difference between $f(x)$ and $\widetilde{f}(x)$ the error, $E$. The aim then is to minimize the $E$ of the $\widetilde{f}(x)$. In physics, we can think of this as attempting to minimize an energy function, while in ML models this is also called minimizing the cost function\cite{moderndeeplearning}. While a loss function is for a single training input, a cost function, on the other hand, is the average loss over the entire training data set. It should be noted that any ML algorithm is an optimization function in itself\cite{harman2012reliable}. This means a hyperparameter optimization scheme in this case can be described as applying the objective function $\widetilde{f}(x)$ on top of the ML objective function $\widetilde{g}(x)$, leading to $\widetilde{f}(\widetilde{g}(x))$.\\

Approaching the hyperparameter optimization for a neural network in such a fashion has two clear advantages. For one, we can apply existing optimization techniques from multiple domains, such as physics in the case of the SA\cite{simulatedanealing} and artificial intelligence in the case of the GA\cite{GeneticSim}, BO\cite{bayesmachine}, and GS\cite{gridsearchcv}. 
The second advantage is that we can split the hyperparameter optimization process from the training process of the ML model. This allows us to use any hyperparameter optimization with any ML technique, as we only apply the hyperparameter optimization to the result of the cost function. Combining these two advantages, we can create a package that can optimize any ML model given a clear cost function.\\

\begin{figure}
\makebox[\linewidth][c]{%
\centering

\includegraphics[width=1\textwidth]{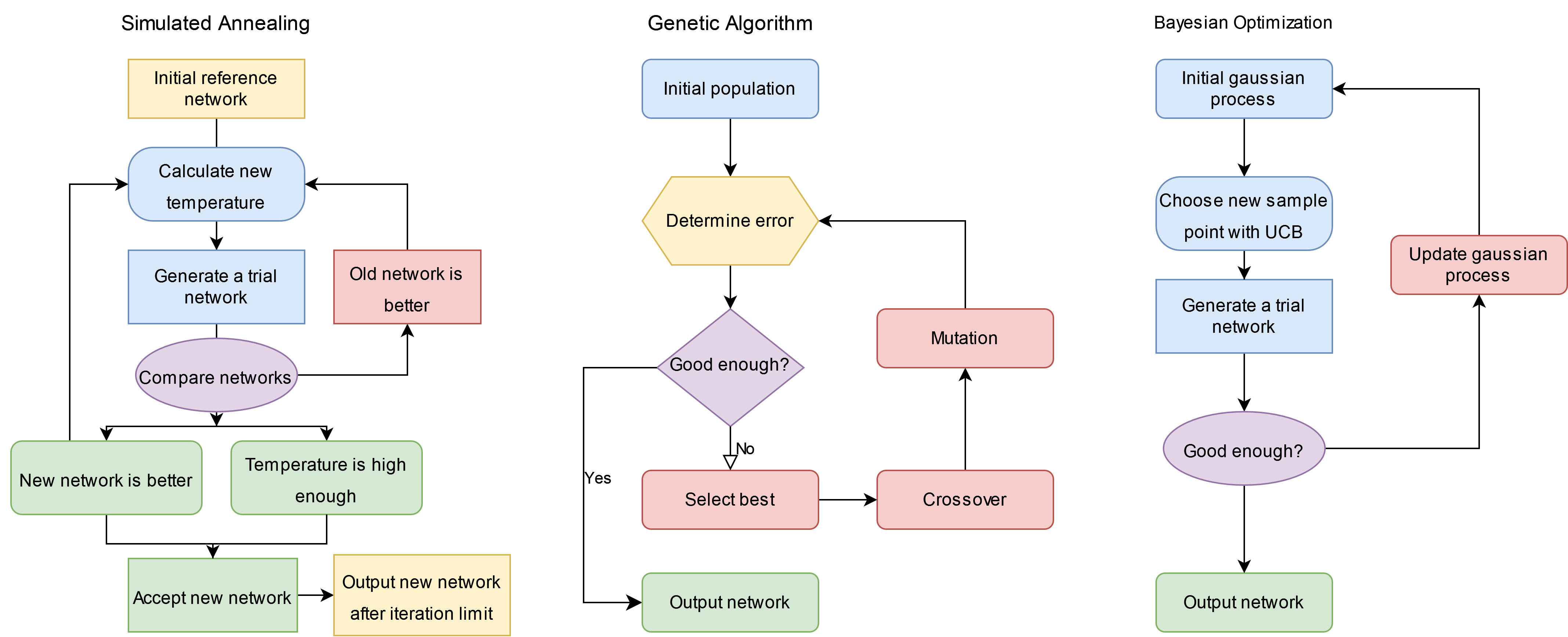}}

\caption{Schematic overview of the SA algorithm (left), the GA algorithm (middle), and the BO algorithm (right).}
\label{fig:Optimizers}
\end{figure}

By implementing various hyperparameter optimization schemes a user is able to generate better end results from the neural network. Currently, HOAX supports  five hyperparameter optimizers, namely the RS(Random Search), GS, SA, GA, and BO.\\

In the RS algorithm, a random position in the hyperparameter space is selected for the neural network to be trained. After the training is completed, the algorithm moves to a new random position. There is no guiding principle behind these moves, other than the type of random function that is called. However, research has shown that the RS can perform equally well in comparison to other methods in some error landscapes and can do so in less computational time\cite{bergstra2012random}. It is therefore a good inclusion as a baseline function.\\

In the GS algorithm\cite{hesterman2010maximum}, the set of $n$ hyperparameters are put in a $n$ dimensional grid. The algorithm defines the grid by adding a step value for each hyperparameter. This step value determines the rate for each hyperparameter change. The GS algorithm is an exhaustive search method. This leads to high computational time for a high number of hyperparameters. However, more advanced options for the GS exist, such as the constraining GS \cite{hesterman2010maximum}, to partially address this issue.\\

In the SA algorithm (see Figure \ref{fig:Optimizers}), which is inspired by the annealing procedure of the metal working \cite{simulatedanealing}, the number of hyperparameters is again divided in a $n$ dimensional grid. However, the algorithm will not visit the entire hyperparameter space in this grid. Therefore, one can choose a larger grid or additional hyperparameters. One starts the algorithm randomly in the hyperparameter space, within which the neural network is created and trained on the training set and is validated with the validation set that gives the minimum error. Then, it moves to a random neighbouring hyperparameter state. For each hyperparameter, it is randomly decided if the parameter is moved or not. If the parameter is moved, the direction is randomly chosen to be either a one-step increase or decrease. Using these new hyperparameters we train a new neural network on the training set that is again validated with the validation set that gives a new minimum error. 
This new state is accepted if the new error, $E_{new}$, is lower than the previous one, $E_{old}$. The temperature, $T$, is a parameter in the SA that affects the distance of a next hyperparameter space from the current state and also the probability of accepting the state with higher objective function value. As the $T$ decreases, the SA reduces the extent of its search to converge to a minimum. There is a probability, $P(A)$, to accept a higher error state depending on the $T$, and the number of iterations $m$, given by equation \ref{Acceptance}.

\begin{equation}
\label{Acceptance}
P(A) = \exp(\frac{-E_{new}-E_{old}}{T/m})
\end{equation}

The number of iterations $m$ determines the chance to accept higher error states, which decreases as $m$ increases. The SA algorithm explores a large search space initially, while gradually reducing the search space to find a minimum. As it is possible that the predefined grid does not include the best hyperparameters, the algorithm also uses an \say{absorbing} boundary\cite{reflectingboundary}. Any time the algorithm reaches one of the hyperparameter boundaries of the grid, it has a certain chance (defined by the user) to remain at the boundary. This is done so that more time can be used to explore the other hyperparameters which have not reached the boundary.\\

In the GA algorithm (see Figure \ref{fig:Optimizers}), which is inspired by the theory of evolution\cite{darwin1859} one uses a \say{gene pool} of six neural networks with random hyperparameters. After training, each of the neural networks is ranked by its error. The two networks with the highest error are removed from the pool. The two with the lowest error become the parent networks for the next generation and are used to generate two new offspring neural networks. To generate the new networks the value of each hyperparameter is converted to a bit-string. A point in this bit-string is selected, called the crossover point. Two new offspring bit-strings are created by swapping part of the parent bit-strings, once before and once after the crossover point for each hyperparameter. For each bit in the bit-string, there is a chance to mutate. The bit will then change from one to zero or vice versa. This rate depends on the bit length of the hyperparameter, which is shown to give improved results\cite{mutationrate,mutation2}. The process is repeated with the new \say{gene pool} until a predefined minimum error or a maximum number of iterations is reached.\\

In the BO algorithm (see Figure \ref{fig:Optimizers}), the bayesian statistics\cite{Norvig}\cite{koch2007bayes} is used to create a probabilistic model able to find the optimal set of hyperparameters. This model is created by sampling points in the hyperparameter space and updating the probabilistic model. The model then uses the statistic distribution to determine the next sampling point. The model consists of two components, the prior function and the acquisition function. The type of prior function determines how the probabilistic model treats uncertainty and the acquisition function determines how the BO decides which point should be sampled next. In this work, we use Gaussian priors and upper confidence bounds (UCB) for the prior function and acquisition functions, respectively \cite{bayesmachine}.

\section{Code infrastructure}
\begin{figure*}

\makebox[\linewidth][c]{%
\centering
\includegraphics[width=1\textwidth]{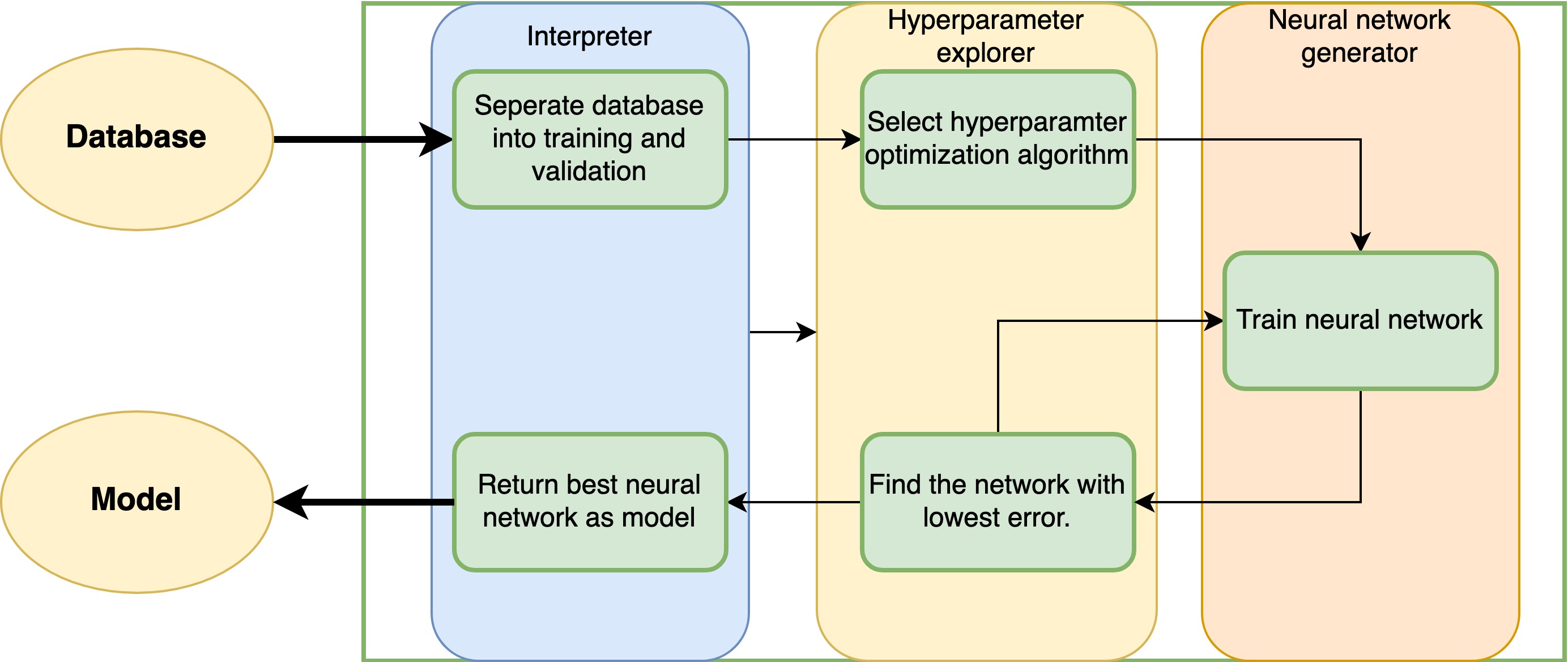}}
\caption{Schematic overview of the HOAX package. The package is divided into three parts; the interpreter, the neural network generator, and the hyperparameter explorer.
}
\label{fig:HOAX}
\end{figure*}

HOAX (hyperparameter optimization algorithm explorer) is an extendable open-source Python package that automates the hyperparameter optimization search for the application of neural network models. HOAX is an extension of the PySurf\cite{pysurf} package, but can also be used as a standalone package. 
The package automates the process of finding hyperparameters which are typically done manually by the user. The user provides a database with training and validation data in the network common data form (NetCDF) or the hierarchical data format version 5 (HDF5)\cite{NetCDF,NetCDF2}, which can be generated by PySurf. The user also provides a configuration file in the JSON (javaScript object notation) format\cite{Json}. This file contains the configuration for processing the database, constructing the neural networks, initializing the fixed hyperparameters, and choosing the hyperparameter optimization algorithm. The package is divided into three parts; i) the interpreter, ii) the neural network generator, and iii) the hyperparameter explorer. A schematic overview of the HOAX can be seen in Figure \ref{fig:HOAX}.\\

The interpreter reads the database that consists of the atom coordinates as input and energies as output. The package contains a module to translate Cartesian coordinates to internal coordinates when this is specified in the configuration file. The interpreter then separates the data into training and validation data sets, according to the user-specified ratio. The training data set is used to train the neural networks, while the validation data is used to measure the performance of the neural network during training.\\ 

The neural network generator initializes the neural networks, which is implemented by providing an interface to the PyTorch library\cite{PyTorch}. Through this interface, the generator creates neural networks based on the specifications provided by the user in the configuration file. The generator trains the neural network with the hyperparameters provided by the hyperparameter optimization algorithm. When a cycle of training has finished, the generator provides a summary of the performance throughout the training, measured using the validation data. It also provides a fully trained neural network.\\

The hyperparameter explorer uses the results provided by the neural network generator to find new hyperparameters. To select a new set of hyperparameters, it uses an optimization algorithm. Currently implemented algorithms are the RS, GS, SA, and BO algorithms. The BO algorithm is implemented using the bayesian optimization python package\cite{bayespackage}. Each of these algorithms can be chosen to provide hyperparameters to the neural network generator. The neural network generator then performs the training sequence again with the new hyperparameters and provides new performance data to the hyperparameter explorer. In this way the process of finding hyperparameters that provides a good fit on the provided data is automated. 
The package also includes a cross-validation module, which can be used to perform X-fold cross-validation on individual hyperparameter configurations found. It can also be used for cross-validation during the training phase while using any of the hyperparameter optimizers. The HOAX package can be found at \url{https://github.com/AlbertThie/HOAX}.

\section{Computational details}

To investigate the effect of the hyperparameter optimizers on neural networks, we used a set of trajectories of \ce{SO2}, Pyrazine (\ce{C4H4N2})\cite{pyrazine}, Furan (\ce{C4H4O}), and Pyrrole (\ce{C4H5N})\cite{PyrFur} as a training data set. This training data set contains the atomic positions of the molecules as training input and the corresponding ground- and two excited-states energies as training output. Training data were created using the PySurf\cite{pysurf} implementation of the Landau–Zener surface hopping simulations\cite{landauzener} to propagate trajectories from initial starting conditions generated by a Wigner sampling\cite{Wigner} based on an initial optimized geometry.\\

For the ab initio calculations, different electronic structure methods were used for the different molecules, to maximize the number of useful data points and minimize the computational cost. All electronic structure calculations were done using the Q-chem 5.4 package\cite{Qchem54}. 
For \ce{SO2}, time-dependent density functional theory (TDDFT) was used with the B3LYP functional\cite{b3lyp} and the 6-31G* basis set\cite{poplegstar}. 
For Pyrazine, we applied TDDFT at the PBE0/cc-pVDZ level of theory.
For Pyrrole and Furan, the spin-flip variant of TDDFT\cite{shao2003spin} was applied using the BHHLYP functional\cite{BHHLYP} and the cc-pVDZ basis set\cite{basisset2}.\\

For each molecule, 100 trajectories were propagated for 100 fs with 0.5 fs timestep. For \ce{SO2} and Pyrazine, the first three singlet states ($S_0$, $S_1$, and $S_2$) were calculated at each point of the trajectory. For Pyrrole and Furan, six states were calculated, and the first three singlet states (the ground and two singlet excited states) were selected based on their spin multiplicity ($S^2$). Each of these calculations yielded a database for the training of at most 20.000 data points. However, not all of the 100 trajectories converged, due to the SCF not converging, and therefore did not make the maximum of 200 data points per trajectory possible for all molecules. For \ce{SO2}, a database of 1541 data points was generated, while for Pyrazine the database contains 20000 data points. For Pyrrole and Furan, 19204 and 14529 data points were generated, respectively. Therefore, not only the fitting and optimization algorithms are tested on different molecular systems, but also for different levels of theories and the different sizes of the training data set.\\

All neural network calculations were performed with the ADAM optimizer\cite{kingma2017adam}, Tanh activation function (see Table \ref{tab2:NNfunctionals}), MSE as error function, and each training session was fixed to last at most 10,000 epochs. During these training sessions, 10\% of training data was not used for training, but used as validation data. This validation data was presented to the network at intervals of 100 epochs. The best-performing iteration of the neural network during the 10,000 epochs was saved and exported by the package as the best-performing model. This was done to prevent overfitting and provide a realistic estimate of the performance of the network on the reference data points (see Figure 3 of the Supporting Information). The reference data points correspond to the coordinates and energies obtained from an independent trajectory with exactly the same initial condition purely based on ab initio calculations without any support of the database and are neither included in the training nor in the validation set. All networks were trained on a data set containing internal coordinates as input and energies as output.

\section{Results and Discussion}

\begin{figure*}

\makebox[\linewidth][c]{
\centering
\subfigure{\includegraphics[width=0.5\textwidth]{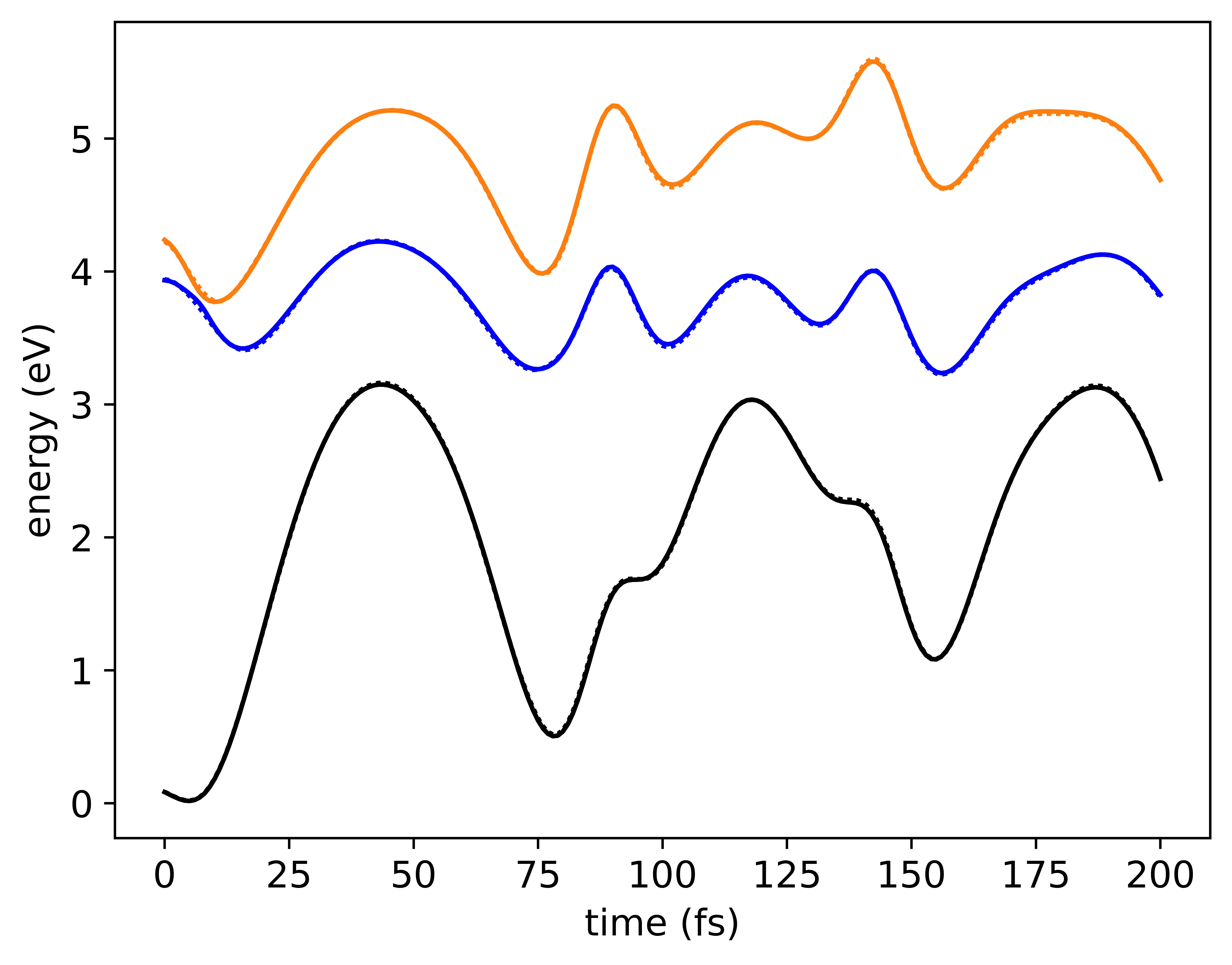}}
\subfigure{\includegraphics[width=0.51\textwidth]{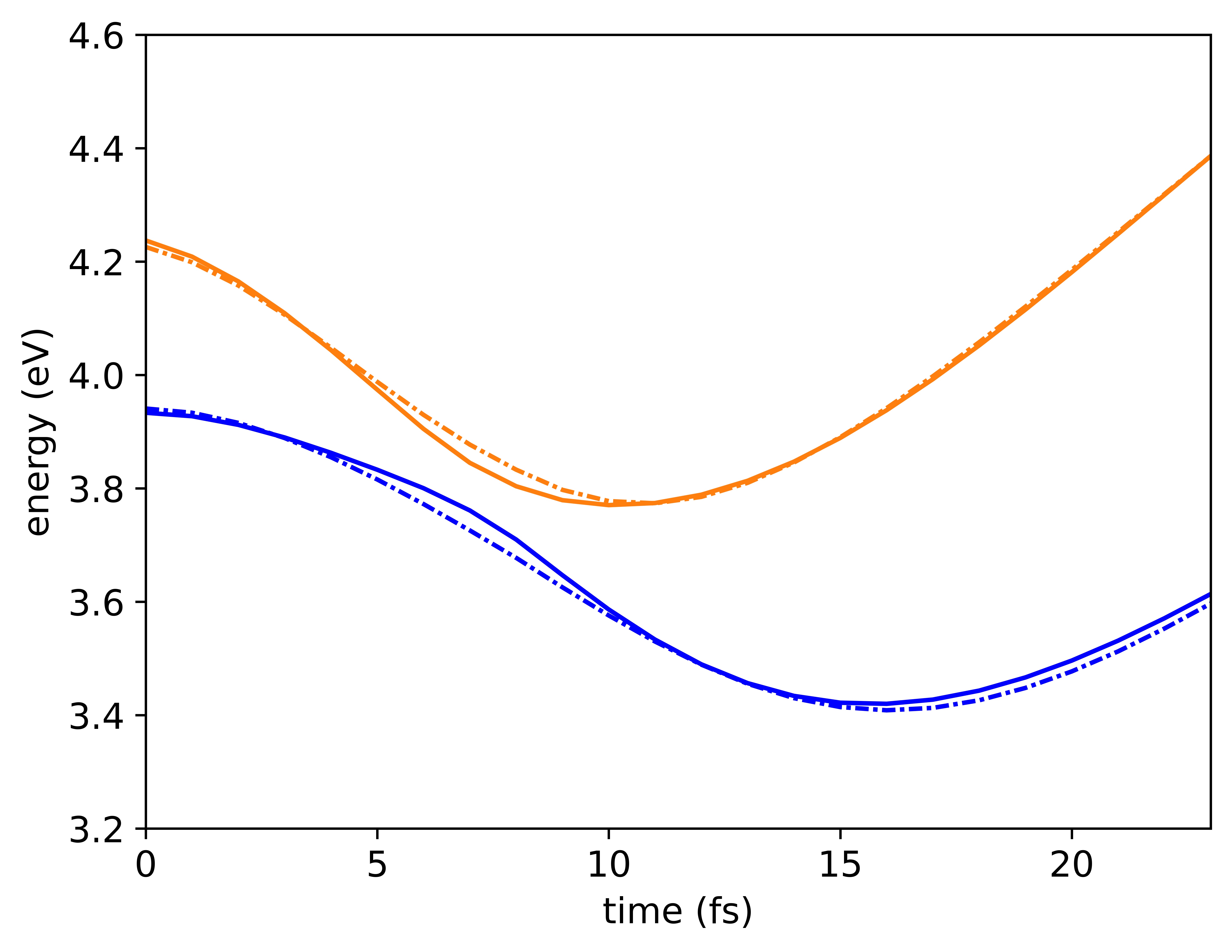}}
}
\caption{Comparison of neural network generated PESs (dashed), for the best network found by the SA algorithm for \ce{SO_2} and reference ab initio based PESs (solid). $S_0$ (black); $S_1$ (blue); $S_2$ (orange). A zoomed-in picture is provided on the right, showing the areas where the $S_1$ and $S_2$ surfaces are close.}
\label{res:SO2}
\end{figure*}

To provide a working prototype, a neural network was trained on the data set of \ce{SO2}, using the SA hyperparameter optimizer. The hyperparameters were initially varied by hand and later explored by the SA optimizer. A network with 150 nodes per layer, nine hidden layers, a learning rate of 0.0001, and a batch size of 32 performed best on the data set. This network produced a RMSE of 0.017 eV on the validation set. The performance compared with the reference trajectory can be seen in Figure \ref{res:SO2}. The results on \ce{SO2} show that the neural network implementation was successful in learning ground- and excited-state PESs. To test the automated cross-validation module, we have performed 10-fold cross-validation on the data set of SO2 using the SA hyperparameter optimizer. The resulting best-performed neural networks, which is composed of 150 nodes per layer, eight hidden layers, a learning rate of 0.0001, and  the batch size of 32, resulted in a RMSE of 0.013 eV on the validation sets. This new neural network architecture differs only in the number of hidden layers compared to the neural networks where no cross-validation was performed  (8 vs. 9 hidden layers), and produces a slightly smaller RMSE (0.013 eV vs. 0.017).\\

To explore the relationship between the performance of the neural network and the hyperparameter optimization techniques, we performed a cross-comparison of the GS, SA, GA, and BO algorithms on Pyrazine, Pyrrole, and Furan. We also investigate the relationship between the molecule that is modelled, the corresponding hyperparameters, as well as the size of the training data.\\

\begin{figure}[H]
    \noindent
    \centering
    \begin{minipage}{.33\textwidth}
        \centering
        \noindent
        \includegraphics[scale=.38,trim={7cm 0 2cm 0}]{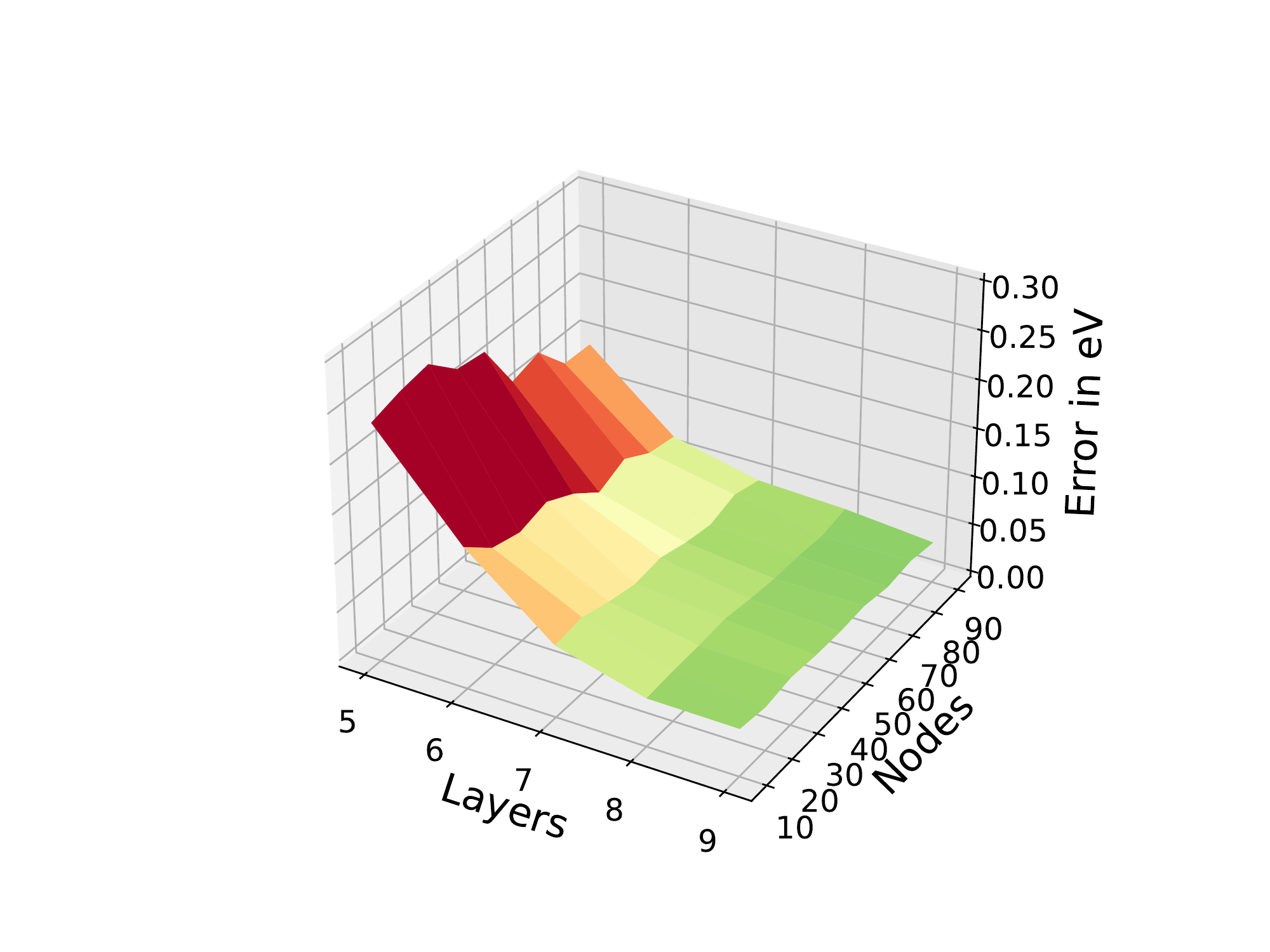}\\
        Pyrazine full data set
    \end{minipage}%
    \begin{minipage}{0.33\textwidth}
        \centering
        \noindent
        \includegraphics[scale=.38,trim={5cm 0 0 0},clip]{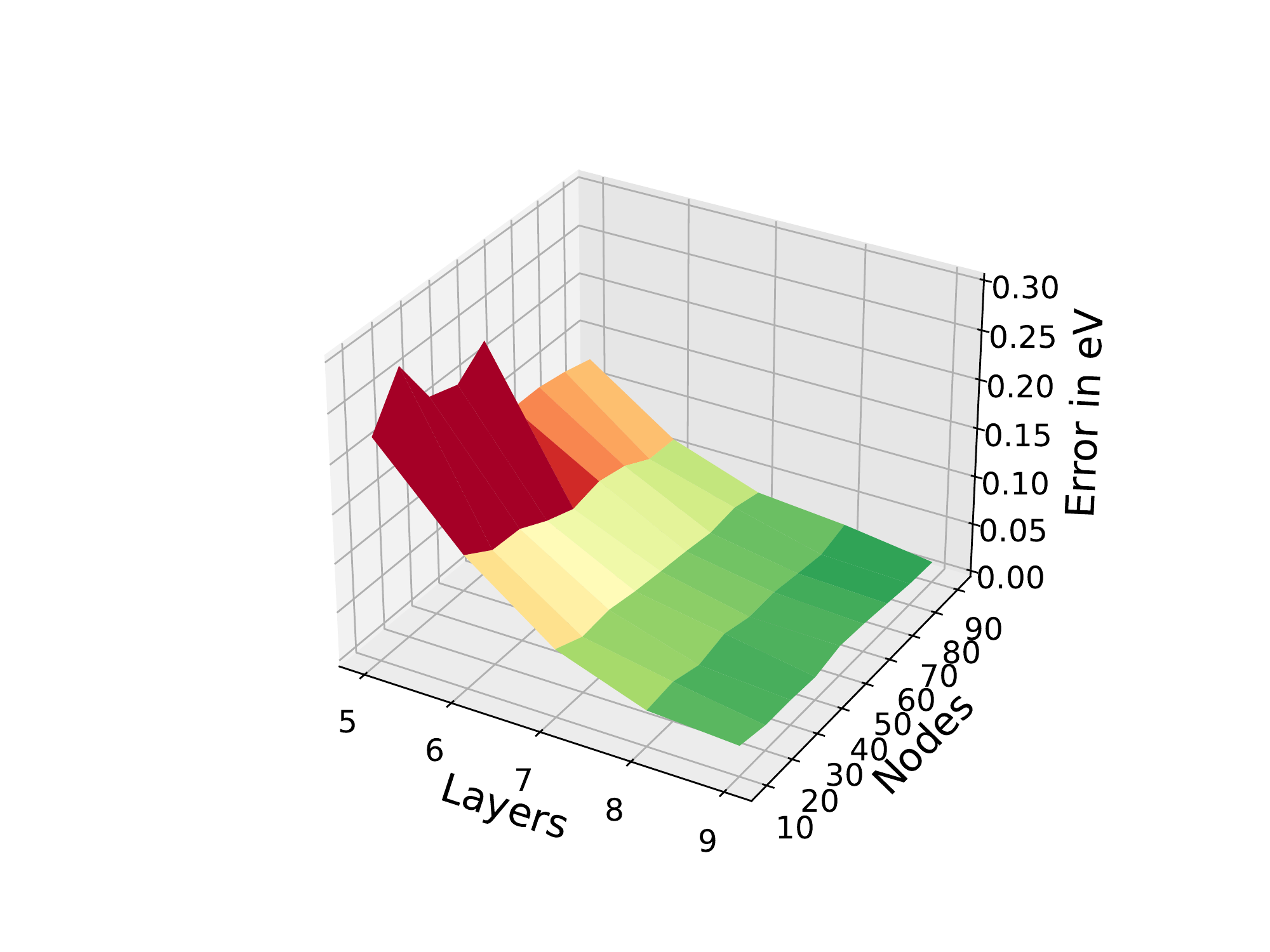}\\
        Pyrrole full data set
    \end{minipage}
    \begin{minipage}{0.33\textwidth}
        \centering
        \noindent
        \includegraphics[scale=.38,trim={5cm 0 0 0},clip]{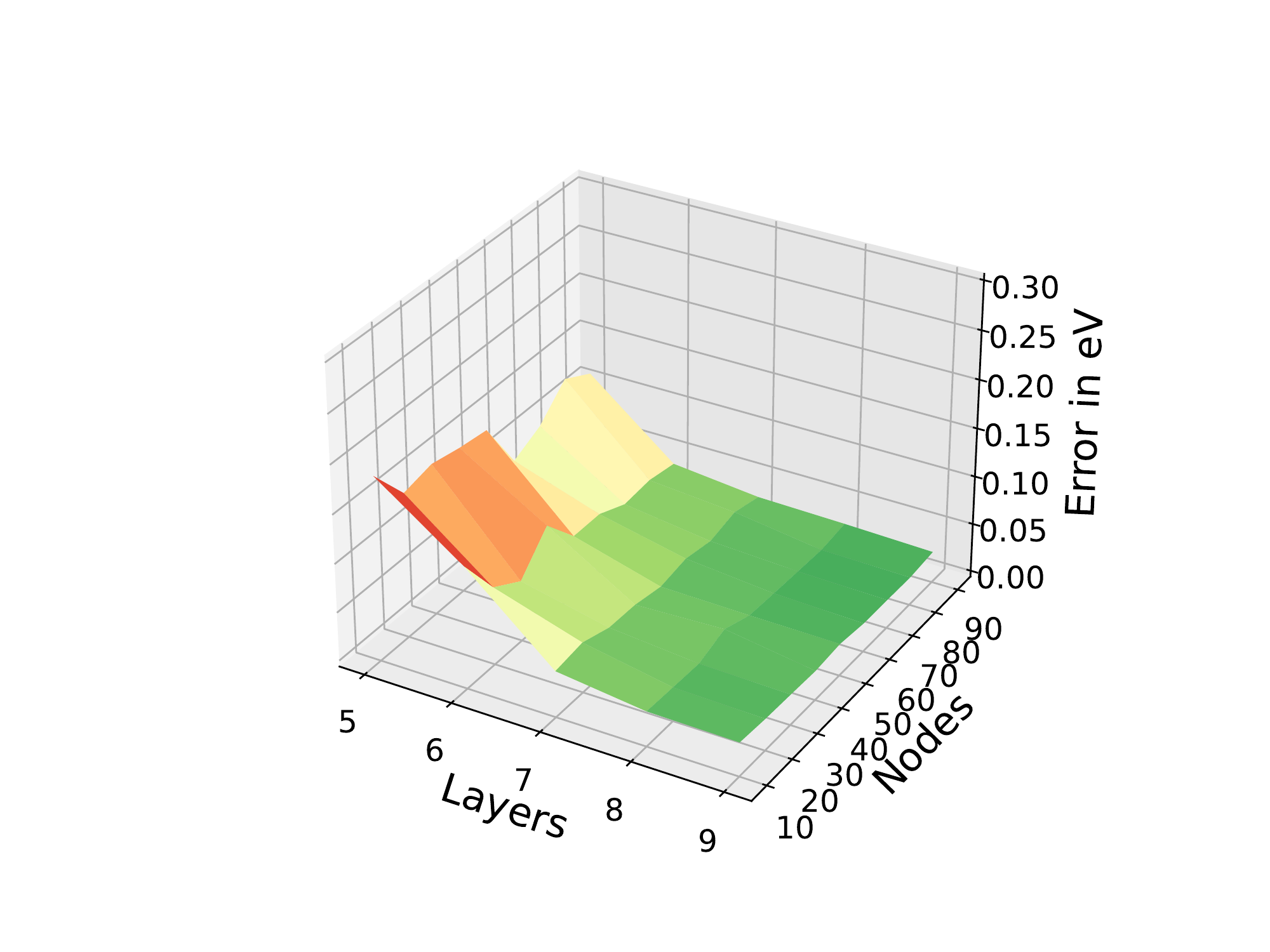}\\
        Furan full data set
    \end{minipage}
    \centering
    \begin{minipage}{.33\textwidth}
        \centering
        \includegraphics[scale=.38,trim={7cm 0 2cm 0}]{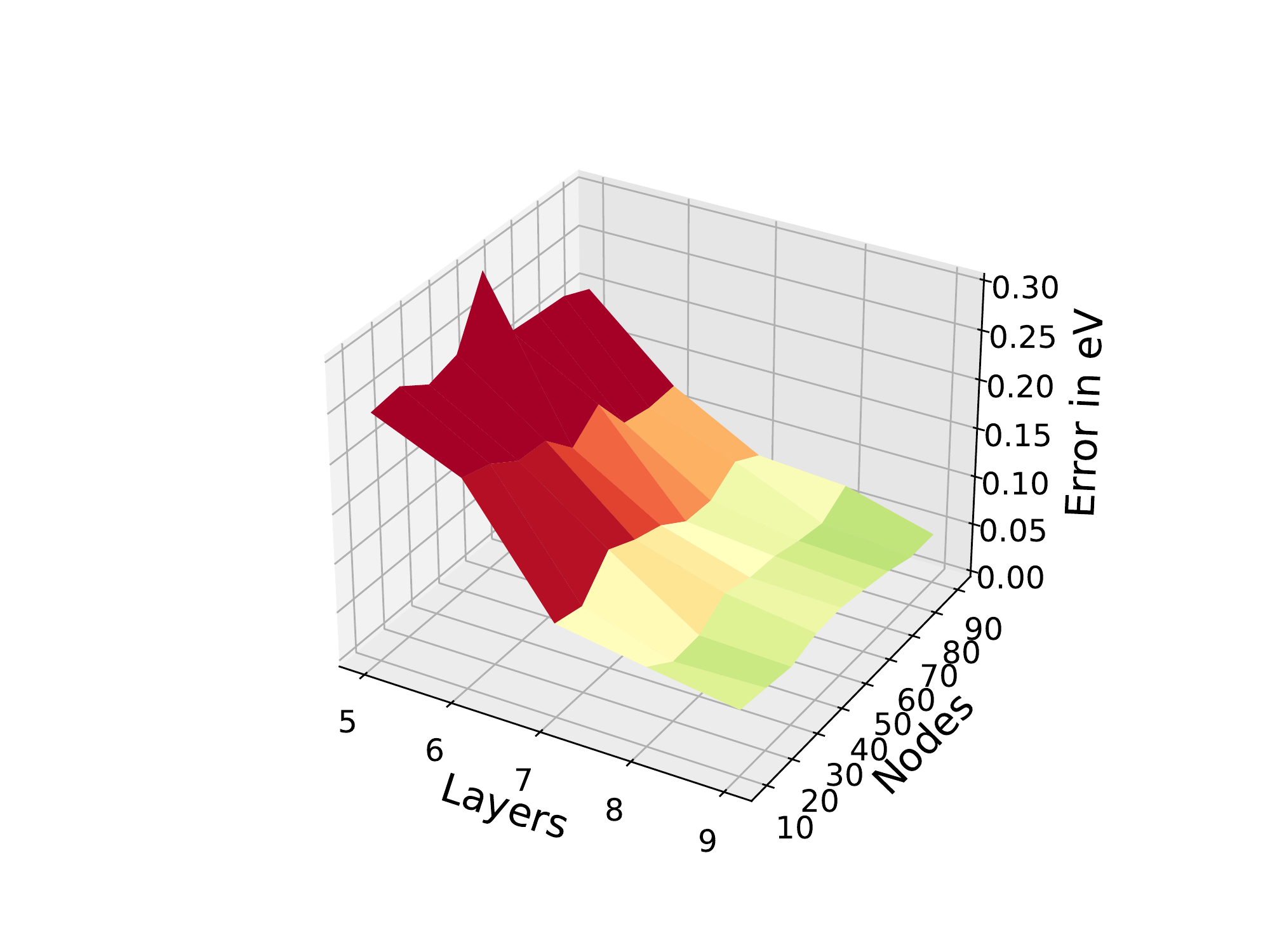}\\
        Pyrazine half data set
    \end{minipage}%
    \begin{minipage}{0.33\textwidth}
        \centering
        \includegraphics[scale=.38,trim={5cm 0 0 0},clip]{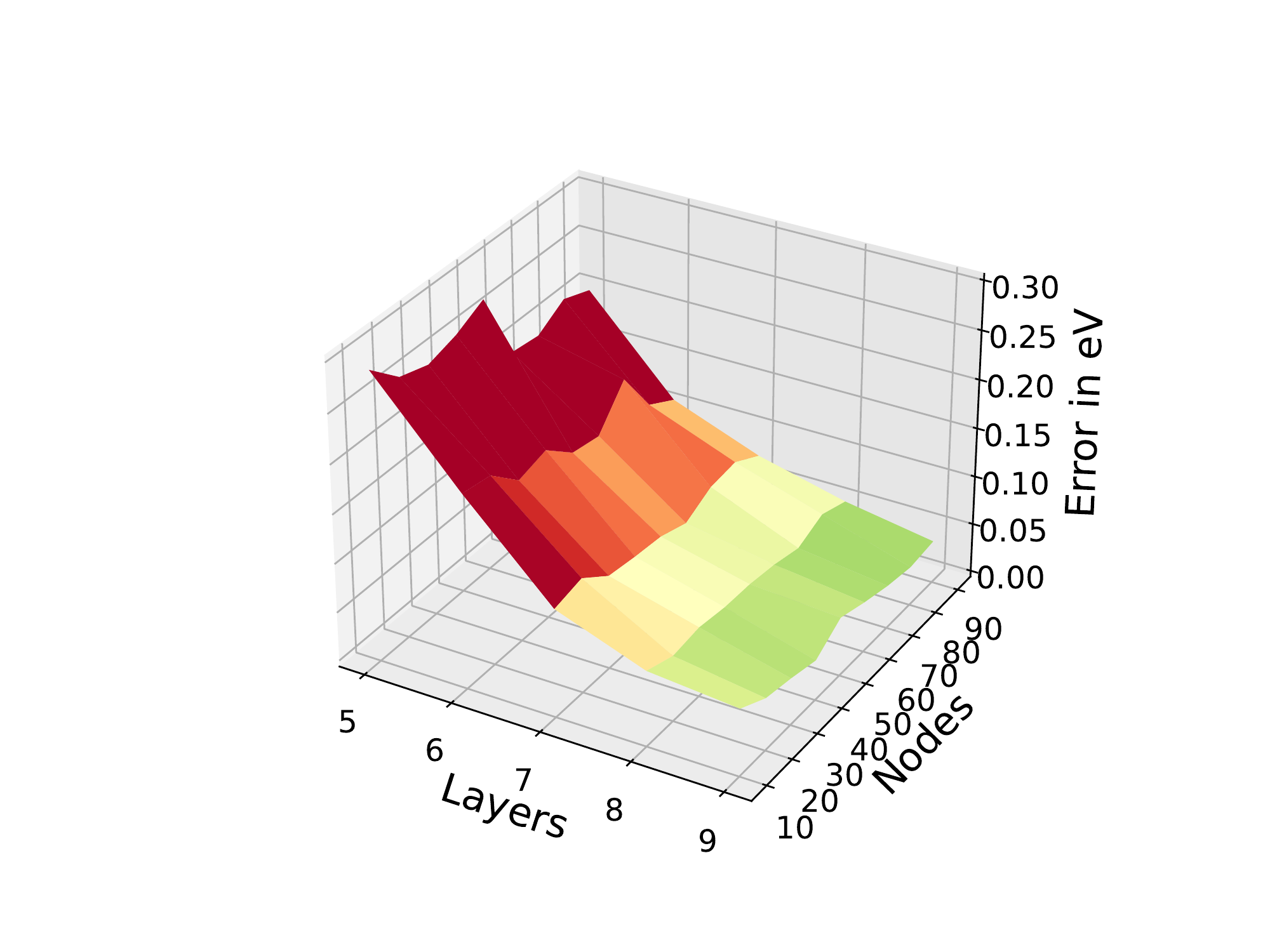}\\
        Pyrrole half data set
    \end{minipage}
    \begin{minipage}{0.33\textwidth}
        \centering
        \includegraphics[scale=.38,trim={5cm 0 0 0},clip]{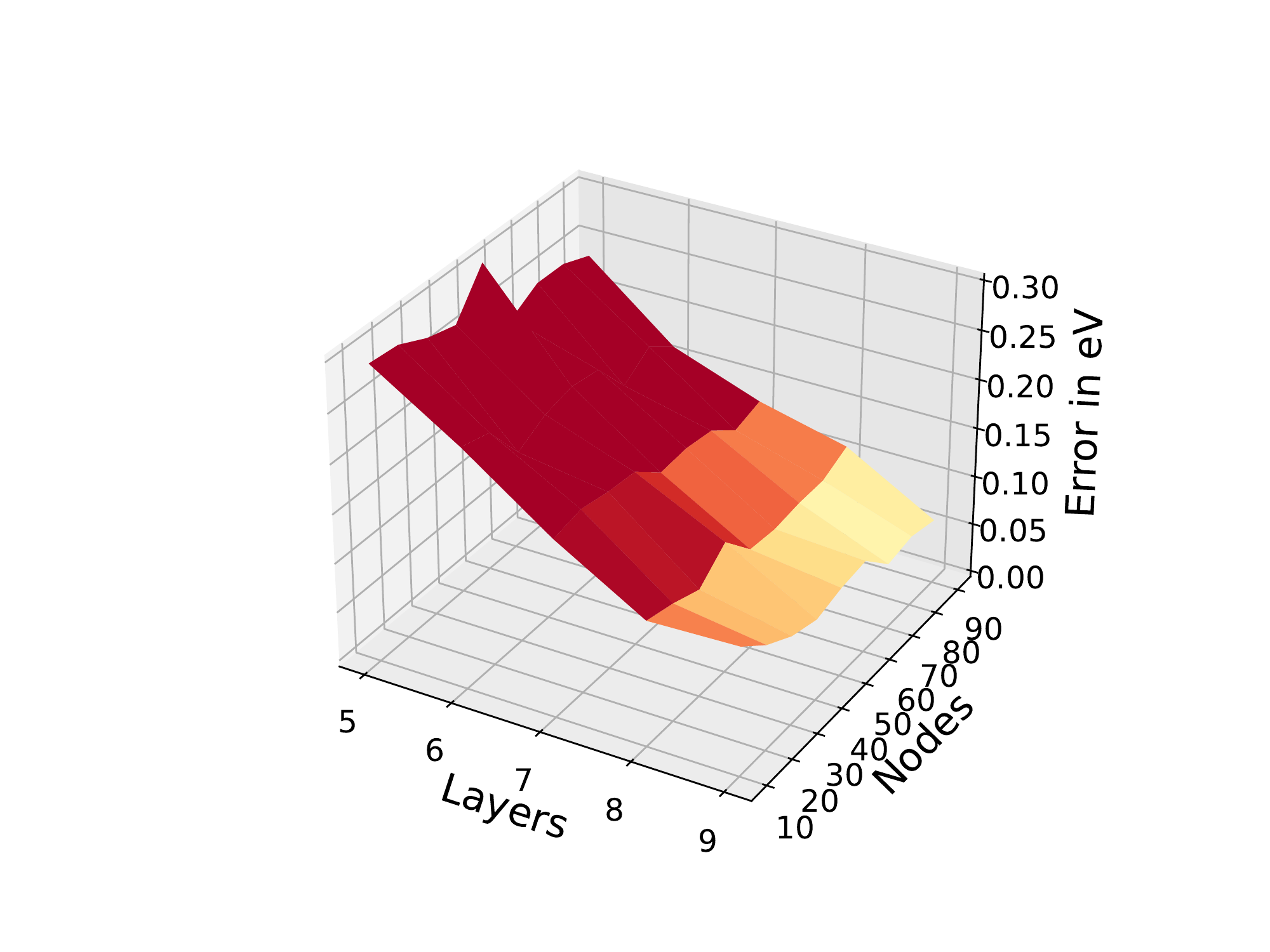}\\
        Furan half data set
    \end{minipage}
\caption{Results of performing the GS between the bounds of 5-9 layers and 10-90 nodes on each molecule. The upper figures show the result for a full data set. The bottom shows the results of using only half the data set. Red coloured areas indicate a high RMSE while green areas indicate a low RMSE.}
\label{fig:gs}
\end{figure}

To study the performance of the GS, the flexible parameters were limited to only two, i.e., the number of nodes and the number of layers. This was done to prevent an exhaustive search of all hyperparameters, which is computationally not feasible. Data was accumulated using the GS with 10 to 90 nodes and 5 to 9 hidden layers. The training was performed with a batch size of 256 and a learning rate of $0.0001$. $10\%$ of the data points were randomly chosen as validation data points and therefore not used in training. The results of the GS algorithm are shown in Figure \ref{fig:gs}. In each point, the lowest error on the validation set is plotted. The slopes of the error descend towards the RMSE of $< 0.025$ eV. The exception is the neural networks trained for Furan on half of the data set. Here the minimum error is $0.034$ eV.
This might be attributed to the low amount of data points available for Furan, leading to a training data set consisting of 6538 data points where the 10\% validation data set is not included. However, the model performs well for the Furan full data set. From these results, we can see that a higher number of layers and nodes has a positive effect on the error, while the effect of the former seems to be more pronounced. 
This can be clearly seen for the Pyrrole full data set results, where a network with 10 nodes and 9 layers can produce an error of $< 0.025$ eV. It can also be seen that the minimum size of the network differs between molecules, where a more complex network is necessary for Pyrazine and Pyrrole compared to Furan. A possible explanation could be that Furan possesses less atoms and therefore it requires less complex modelling. Another explanation could be that the initial parameters selected for the GS algorithm are closer to the optimal parameters for Furan.

Extending the GS optimizer beyond node and layer parameters is computationally costly, as the cost increases exponentially with respect to the number of hyperparameters. To this end, and to avoid a full scan of the hyperparameter surface, the metaheuristic algorithms SA, GA, and BO are used\cite{simulatedanealing,GeneticSim,bayesmachine}. Using these hyperparameter optimizers, the parameter search space is increased without increasing the computational costs. This allowed us to include four hyperparameters, i.e., the number of layers and nodes, learning rate, and batch size, as presented in Tables \ref{tab3:SA}, \ref{tab4:GA} and \ref{tab5:BO}.

\begin{table}[H]

\small
\caption{Hyperparameter configurations for the simulated annealing optimizer. Each column shows the lowest error found on the data set and the hyperparameters used in the model.\\}
\begin{tabular}{ |p{3.0cm}|p{1.5cm}|p{1.5cm}|p{1.5cm}|p{1.5cm}|p{1.5cm}|p{1.5cm}|}
 \hline
 \multicolumn{7}{|c|}{Simulated Annealing Results} \\
 \hline
 \hline
   & Pyrazine & Pyrazine Half & Furan & Furan Half & Pyrrole & Pyrrole Half\\
 \hline
 Number of Nodes & 120   & 60 &   140 &150&60&40\\
 Number of Layers&  4 & 5   & 19 &13&6&11\\
 Learning Rate &0.0001 & 0.0001&  0.0001&0.0001&0.0001&0.0001\\
 Batch Size   & 64 & 32 &  128 &128& 32&32\\
 RMSE in eV & 0.023 & 0.029 &0.020& 0.015 &0.016&0.029\\
MAE in eV &0.013 & 0.015 & 0.012& 0.010 & 0.010& 0.015\\
 \hline
\end{tabular}
\label{tab3:SA}
\end{table}
\begin{table}[H]

\small
\caption{Hyperparameter configurations for genetic algorithm optimizer. Each column shows the lowest error found on the data set and the hyperparameters used in the model. \\}
\begin{tabular}{ |p{3.0cm}|p{1.5cm}|p{1.5cm}|p{1.5cm}|p{1.5cm}|p{1.5cm}|p{1.5cm}|  }
 \hline
 \multicolumn{7}{|c|}{Genetic Algorithm Results} \\
  \hline
 \hline
   & Pyrazine & Pyrazine Half & Furan & Furan Half & Pyrrole & Pyrrole Half\\
 \hline
 Number of Nodes & 80   & 100 &   180 &180&70&160\\
 Number of Layers&  12 & 18   & 17 &11&14&16\\
 Learning Rate &0.0001 & 0.0001&  0.0001&0.0001&0.001&0.0001\\
 Batch Size   & 256 & 128 &  256 &256& 256&32\\
 RMSE in eV & 0.027 & 0.016 &0.019& 0.016 &0.018&0.013\\
  MAE in eV &0.015& 0.010& 0.012 &0.011  &0.012 &0.010\\
 \hline
\end{tabular}
\label{tab4:GA}
\end{table}
\begin{table}[H]
\label{tab5:BO}
\small
\caption{Hyperparameter Configurations for the bayesian optimizer. Each column shows the lowest error found on the data set and the hyperparameters used in the model.\\}
\begin{tabular}{ |p{3.0cm}|p{1.5cm}|p{1.5cm}|p{1.5cm}|p{1.5cm}|p{1.5cm}|p{1.5cm}|  }
 \hline
 \multicolumn{7}{|c|}{Bayesian Optimization Results} \\
  \hline
 \hline
   & Pyrazine & Pyrazine Half & Furan & Furan Half & Pyrrole & Pyrrole Half\\
 \hline
 Number of Nodes & 90   & 190 &   60 &190&80&130\\
 Number of Layers&  14 & 6   & 13 &5&7&5\\
 Learning Rate &0.0001 & 0.0001&  0.0001&0.0001&0.0001&0.0001\\
 Batch Size   & 128 & 128 &  32 &32& 32&64\\
 RMSE in eV & 0.021 & 0.018 &0.021& 0.029 &0.017&0.016\\
MAE in eV &0.014 &0.013  & 0.011&0.014 & 0.010& 0.011\\
 \hline
\end{tabular}
\end{table}

The SA and BO results in Table \ref{tab3:SA} and \ref{tab5:BO} were reached after 20-30 iterations, while the results for the GA in Table \ref{tab4:GA} were taken after 10-15 iterations. The slightly lower iterations for the GA is attributed to the fact that the GA will generate two new networks at each iteration, while SA and BO will generate one. On average the required computational time between the SA, GA, and BO is comparable, as the computational time depends on the training of the neural network. The computational time also depends on the hyperparameters of the neural network and the number of data points. A larger network increases computational time, while using a halved data set reduces the computational time. For all optimizers, 144 hours of training on 4 Intel(R) Xeon(R) CPU E5-2695(v4) with 2.10GHz generated the results. All optimizers outperformed the GS in reaching a lower RMSE. All optimizers need a different configuration of hyperparameters to reach the lowest RMSE. This could indicate that the error landscape contains many local minima. It is also the case that the hyperparameters themselves are correlated. For example, the learning rate found for the neural networks for Pyrrole by the GA in Table \ref{tab4:GA} is higher than the one found by SA in Table \ref{tab3:SA}. However, the batch size is much higher, meaning the learning updates are performed much less frequently than in the SA algorithm. In a similar fashion, the Furan model found by the SA has a higher number of layers but a smaller number of nodes than the one found by the GA. We can therefore see that for different molecules, totally different hyperparameters are necessary to reach a well-performing network. This is in fact related to the dependency of the hyperparameters to the target property for the neural network methods and acts as a limiting factor in the transferability of the optimal hyperparameters to new molecules. However, finding similarities for the bounds of the hyperparameter search area for different molecules helps in narrowing these search bounds and thus saves time in finding the best hyperparameters for new molecules with comparable size and complexity. We can however see a trend where networks with more layers combined with a large batch size are more likely to produce more favorable results, whereas smaller networks require a smaller batch size to be successful. These relations reflect the complexity of finding an optimal set of hyperparameters.

In addition, the hyperparameters are in most cases dependent on each other, and therefore the resulting high dimensional error surface that the optimizers have to navigate through shows many local minima. It is true that for fully converged optimizations it is expected that all three optimizers (GA, SA, and BO) lead to similar results, however, finding these global minima is a challenge itself and is beyond the scope of this work. In this work, we focus on finding the optimized hyperparameters leading to an error lower than a given threshold. The configurational hyperparameter space that the SA, GA, and BO optimizer visit during the optimization is given by their respective algorithm and can be significantly different from each other. This leads in our opinion to the discrepancy between the optimal found hyperparameters given the different optimizers.

We can also see that halving the number of data points results in different hyperparameters, but still workable neural networks. This shows that the optimizer is capable of producing workable hyperparameters for different amounts of data points. It should be noted that the poor performance of the GS that used two hyperparameters for the Furan half data set is no longer the case when the SA, GA, and BO algorithms are used with four hyperparameters.\\

\begin{figure}[!htb]
\makebox[\linewidth][c]{%
\centering
\subfigure{\includegraphics[width=0.35\textwidth]{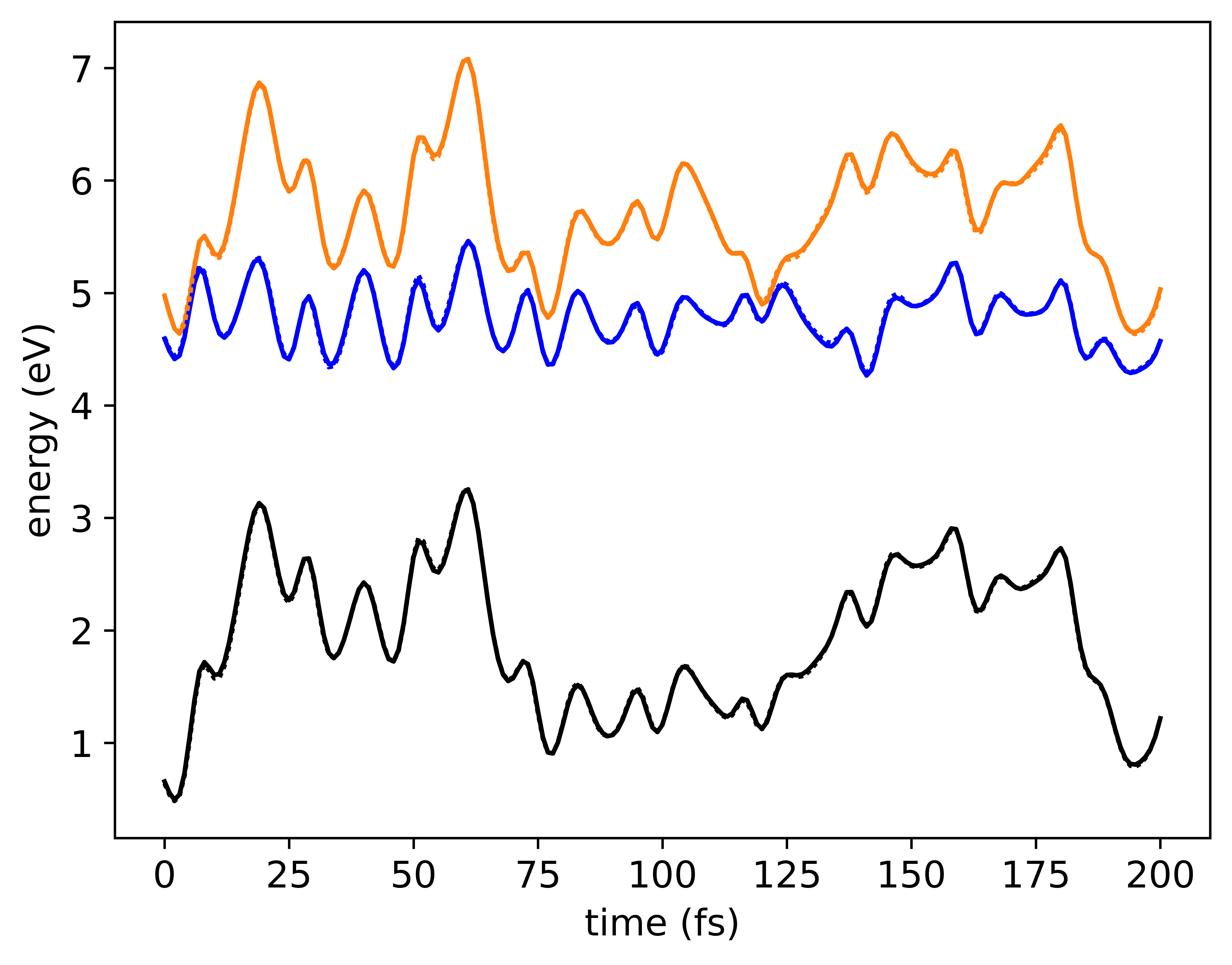}}%
\subfigure{\includegraphics[width=0.35\textwidth]{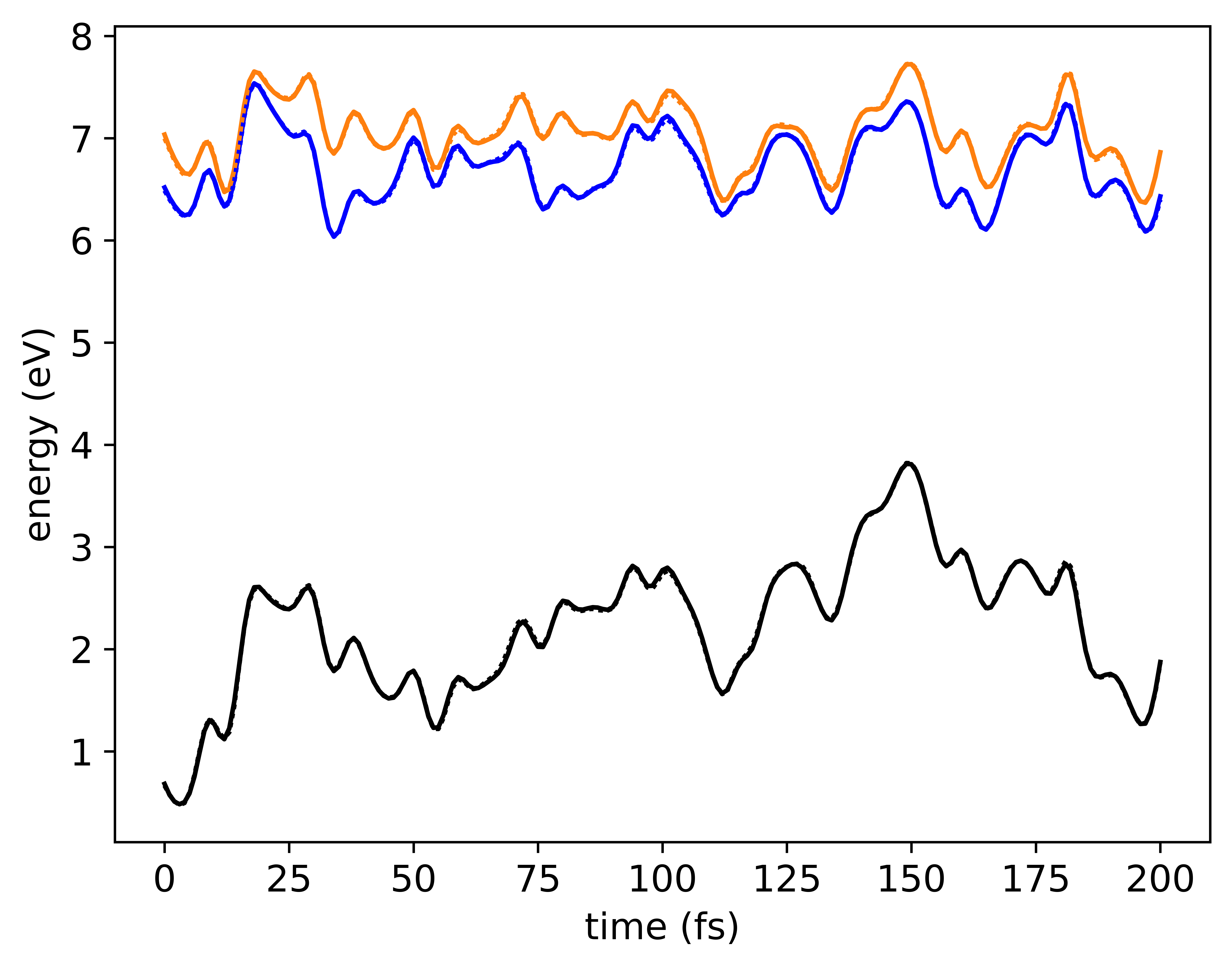}}%
\subfigure{\includegraphics[width=0.35\textwidth]{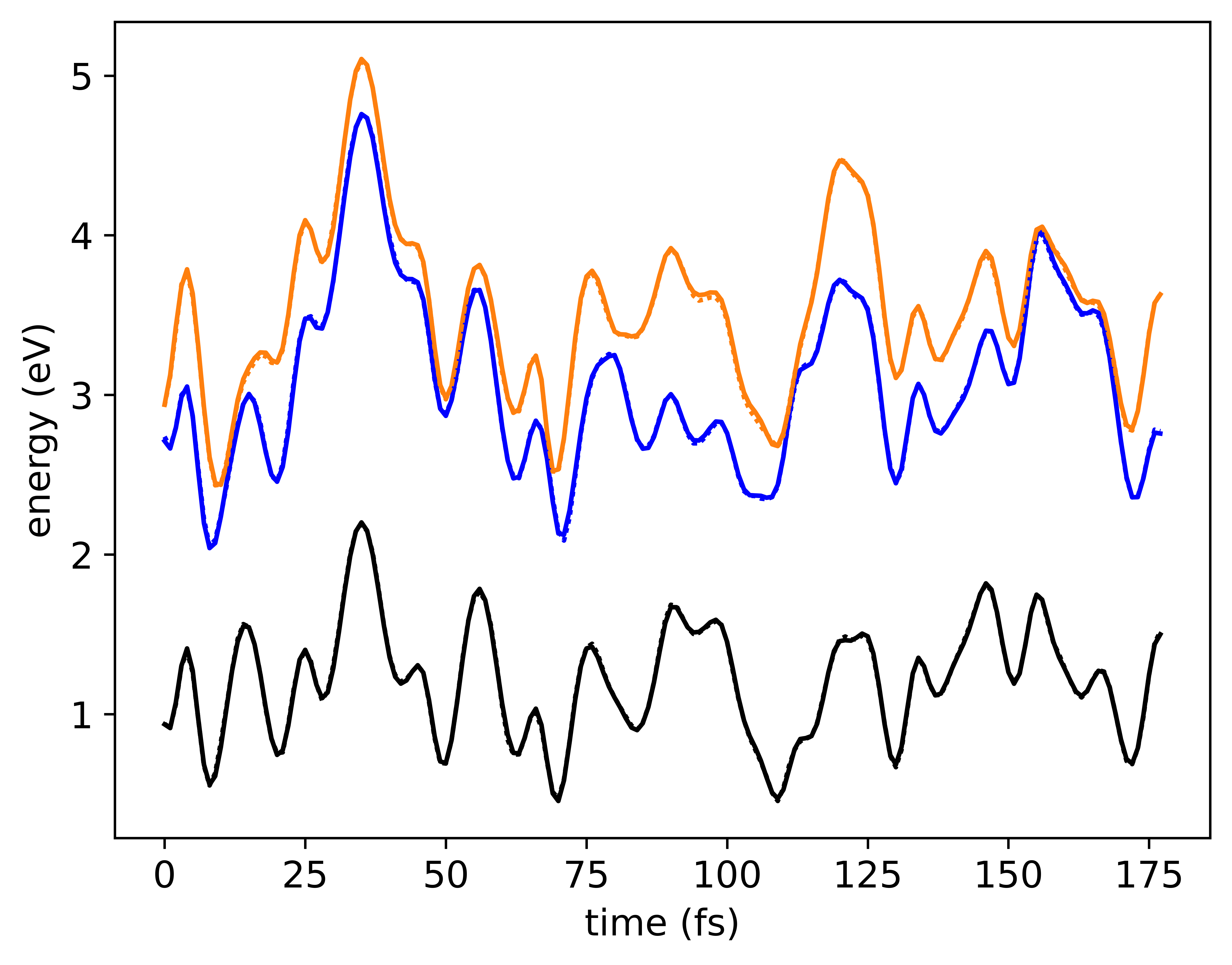}}%
}
\caption{
Comparison of neural network generated PESs (dashed), for the best network found by the SA algorithm for Pyrazine (left), the GA algorithm for Furan (middle), and the BO algorithm for Pyrrole (right) and reference ab initio based PESs (solid). $S_0$ (black); $S_1$ (blue); $S_2$ (orange).}
\label{fig:trajectoriesopt}
\end{figure}

It is important to check the validity and reliability of the generated neural network-based PESs that are prerequisites for sound non-adiabatic excited-state dynamics such as surface hopping simulations. To verify this, in Figure \ref{fig:trajectoriesopt} we show the PESs as functions of time for the reference ab initio based trajectory (solid) together with the neural network-generated PESs constructed with the best hyperparameters 
(dashed). It is hard to see any difference as the curves lie on top of each other, reflecting their excellent agreement. The neural network-generated PESs are smooth and confirm the shape of the reference trajectories. The model PESs are able to capture the different characteristics for each molecule, e.g., the higher energy gap of Furan and the steeper energy landscape of Pyrrole.
\begin{figure}[!htb]
\makebox[\linewidth][c]{%
\centering
\subfigure{\includegraphics[width=0.35\textwidth]{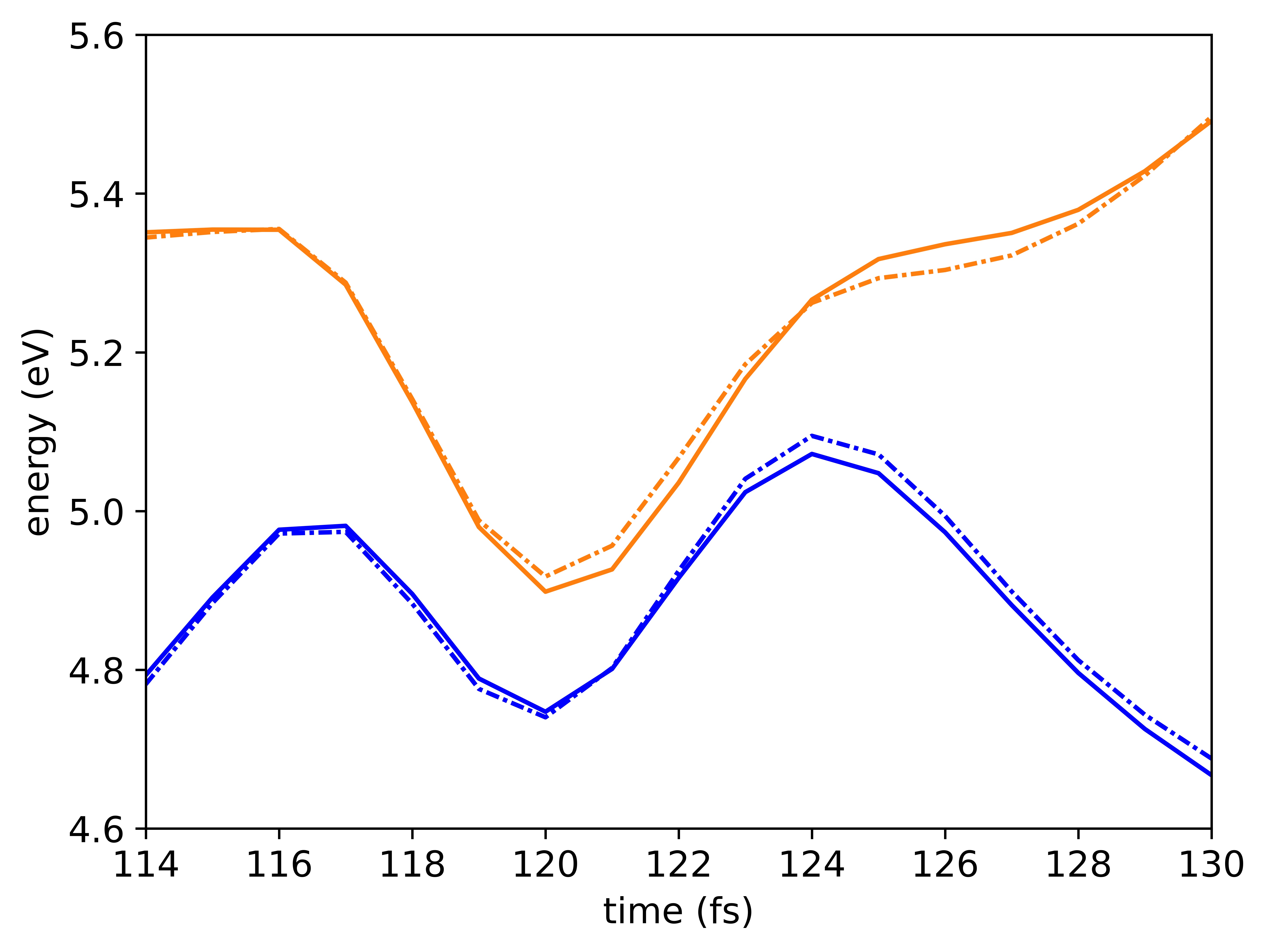}}
\subfigure{\includegraphics[width=0.35\textwidth]{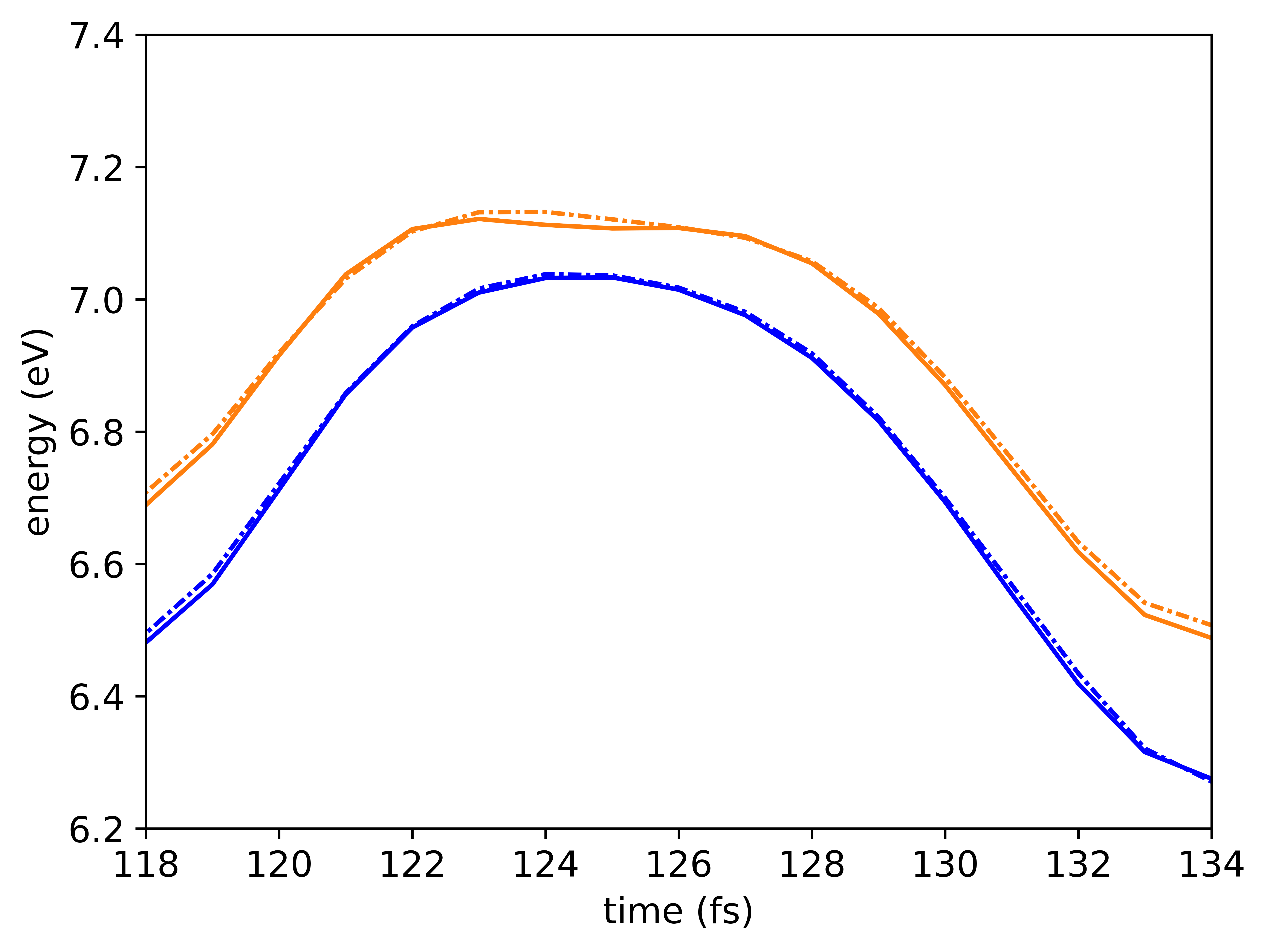}}%
\subfigure{\includegraphics[width=0.35\textwidth]
{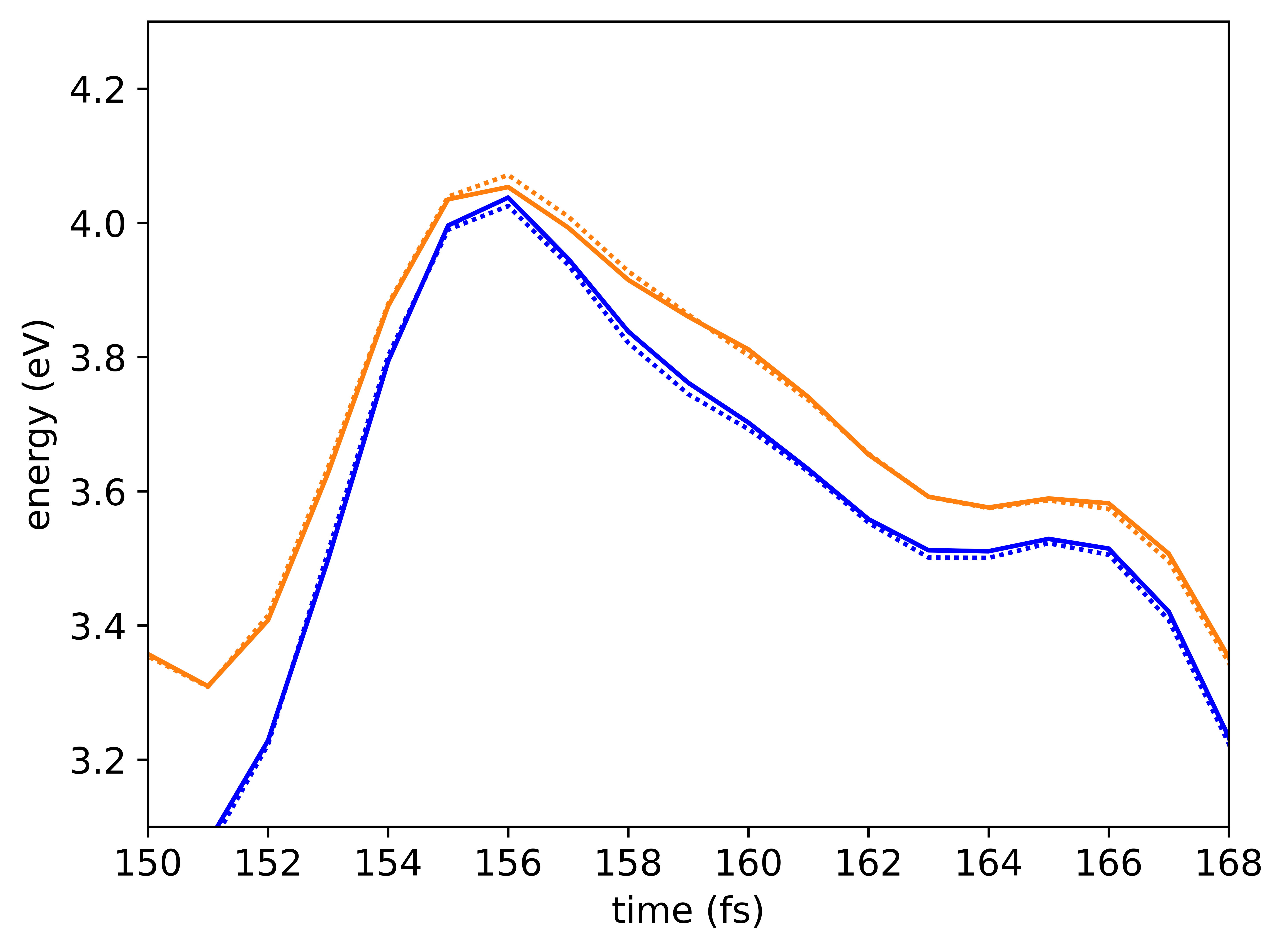}}%
}
\caption{Comparison of neural network generated PESs (dashed), for the best network found by the SA algorithm for Pyrazine (left), the GA algorithm for Furan (middle), and the BO algorithm for Pyrrole (right) and and reference ab initio based PESs (solid), zoomed in on the areas where the $S_1$ and $S_2$ surfaces are close. $S_1$ (blue); $S_2$ (orange).}
\label{fig:trajectorieszoomopt}
\end{figure}

A good model PESs relies on appropriate training data set that naturally sample the important parts of the conformational space including the areas of interest to surface hopping, namely the areas of high hopping probabilities that is when the PESs are getting close to each other. Figure \ref{fig:trajectorieszoomopt} shows a close view of the crucial areas, i.e. where the $S_1$ and $S_2$ surfaces are close to each other. It is clear that these important areas are also modelled well. The neural network-generated PESs (dashed) show a slightly larger energy gap than the reference PESs (solid). The difference in the energy gap is about 11 meV, which is significantly small considering the fact that neural network-generated PESs require less human effort and reduce much of the computational cost associated with constructing the reference PESs. Analogous Figures for the best networks found by the GR for Pyrazine, Pyrrole, and Furan can be found in the Supporting Information (Figure 1-2).

\section{Conclusion}

The HOAX package is an extendable open-source package written fully in Python that facilitate the hyperparameter optimization search for the
application of neural network models for constructing PESs in an automated fashion, which bypasses the need for a lengthy
manual process and reduces computational costs compared to the ab initio based PESs. It uses the PyTorch interface to generate fast and customized implementations of neural networks. Additionally, it is currently an extension of the PySurf package as a new Plugin engine. Thus, it has direct access to the PySurf database for training and validation data in the NetCDF and the HDF5 formats. HOAX can be easily adapted to other packages that provide molecular data in HDF5 compatible format. The hyperparameter optimizer could also be extended to perform automated cross-validation on the best networks found, even during the neural network training process.

In this work we show that the package can produce neural network model PESs for \ce{SO2}, Pyrazine, Pyrrole, and Furan, using the GS, SA, GA, and BO algorithms. The GS algorithm results show that a higher number of layers and nodes has a positive effect on the error. The BO, SA, and GA models show similar performance at the same computational time. All three hyperparameter optimizers are able to search a four-dimensional hyperparameter space and consistently find the optimal hyperparameters configuration. The 50\% reduction of the data set still produces good model PESs, showing the robustness of the model.  The neural network-generated PESs are smooth and confirm the shape of the reference ab initio based trajectories. The model PESs are able to capture the different characteristics for each molecule. The areas of interest to surface hopping, namely the areas of high hopping probabilities that is when the PESs are getting close to each other, are also modelled well.\\

The HOAX package currently supports the optimization of neural network hyperparameters, but it can be extended to other ML models such as,  kernel-based methods\cite{kernelLilienfeld2015} , Gaussian processes\cite{Srinivas2012}, and random forest regression\cite{randomforest}.
Also, different neural network architectures could be introduced, such as graph neural networks\cite{graph1} or convolutional neural networks\cite{albawi2017understandingconvol}.
The hyperparameter optimizers can also be extended with additional non-gradient methods, such as a greedy randomized adaptive search procedure\cite{feo1995greedy}. Different molecular representations for the training data, such as Coulomb-matrices or graph representations could be considered. The modular setup of the package provides flexibility to add custom functionalities. Work in this direction is currently in progress in our research team. The model can also be extended for other types of data, such as charges, dipole moments, and orbital energies, as long as the data is provided in the HDF5 or NetCDF format. 

\section{Acknowledgement}
Prof. Peter Gill is one of the main contributors to the electronic structure program Q-Chem and drove the development of theoretical chemistry in his successful research carrier. We are happy to honor him with this debut of the HOAX software package. The authors thank Prof. Peter Gill, Dr. Andrew Gilbert, and E. Salazar for many inspiring and intellectually stimulating conversations and for giving an insight into the process of new method development. Additionally, the authors are thankful to Dr. Kiana Moghaddam for her help in proofreading the manuscript.
This work is part of the Innovational Research Incentives
Scheme Vidi 2017 with project number 016.Vidi.189.044,
which is financed by the Dutch Research Council
(NWO).

\bibliographystyle{tfo}
\bibliography{bibliography.bib}

\end{document}


\author{
\name{Albert Thie\textsuperscript{a}, Maximilian F.S.J. Menger\textsuperscript{a} and Shirin Faraji*\textsuperscript{a}\thanks{CONTACT Shirin Faraji  Email: s.s.faraji.rug.nl}}
\affil{\textsuperscript{a}Zernike Institute for Advanced Materials, Faculty of Science and Engineering, University of Groningen, Nijenborgh 4, 9747AG Groningen The Netherlands. }
}
\title{Supplementary Information - HOAX:  A Hyperparameter Optimization Algorithm Explorer for Neural Networks}
\maketitle

\begin{keywords}
Quantum Chemistry, Machine Learning, Neural Networks, Hyperparameter Optimization
\end{keywords}



\subsection{Grid Search Trajectories}

\begin{figure}[H]
\makebox[\linewidth][c]{%
\centering
\subfigure{\label{fig:aa}\includegraphics[width=0.35\textwidth]{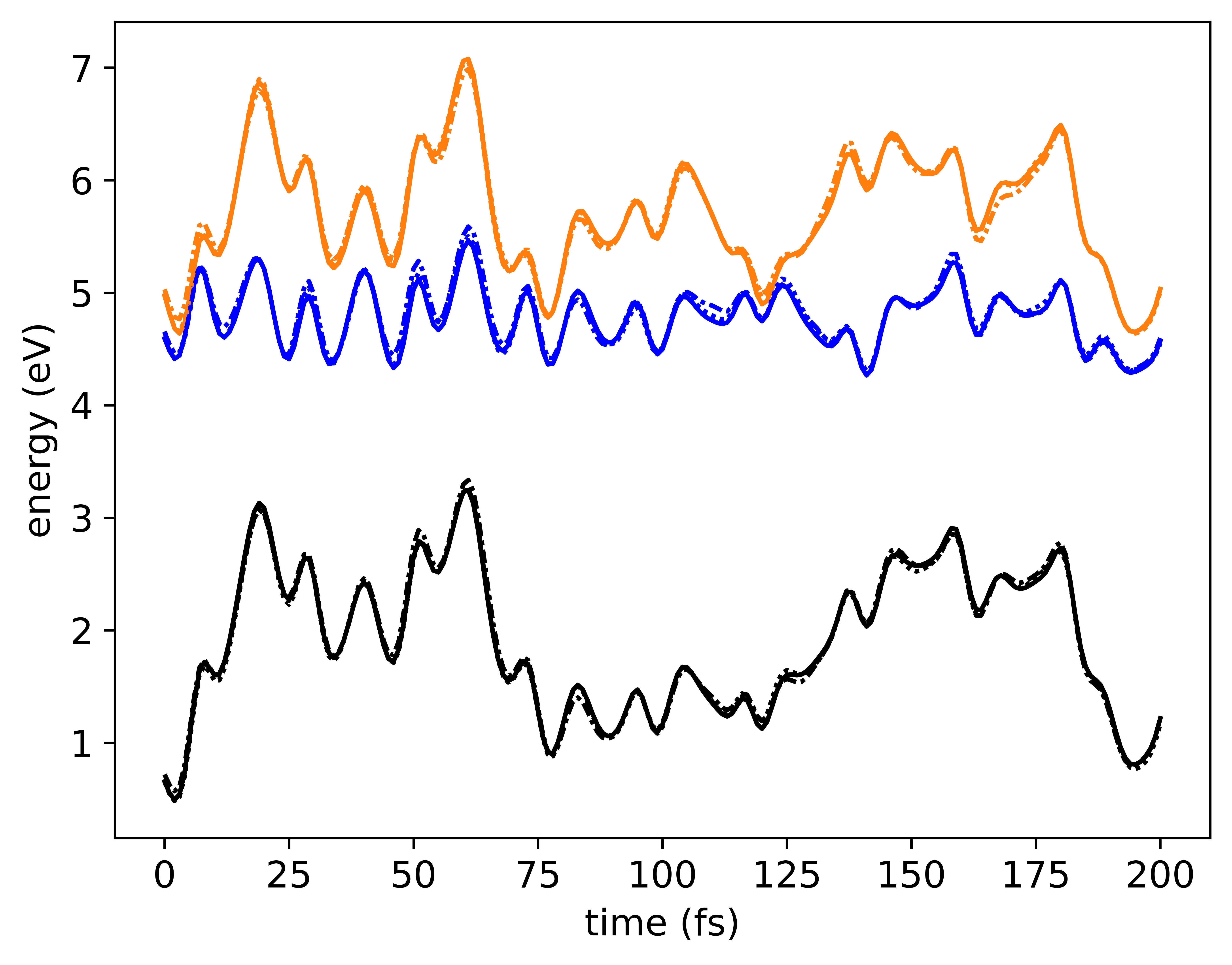}}%
\subfigure{\label{fig:bb}\includegraphics[width=0.35\textwidth]{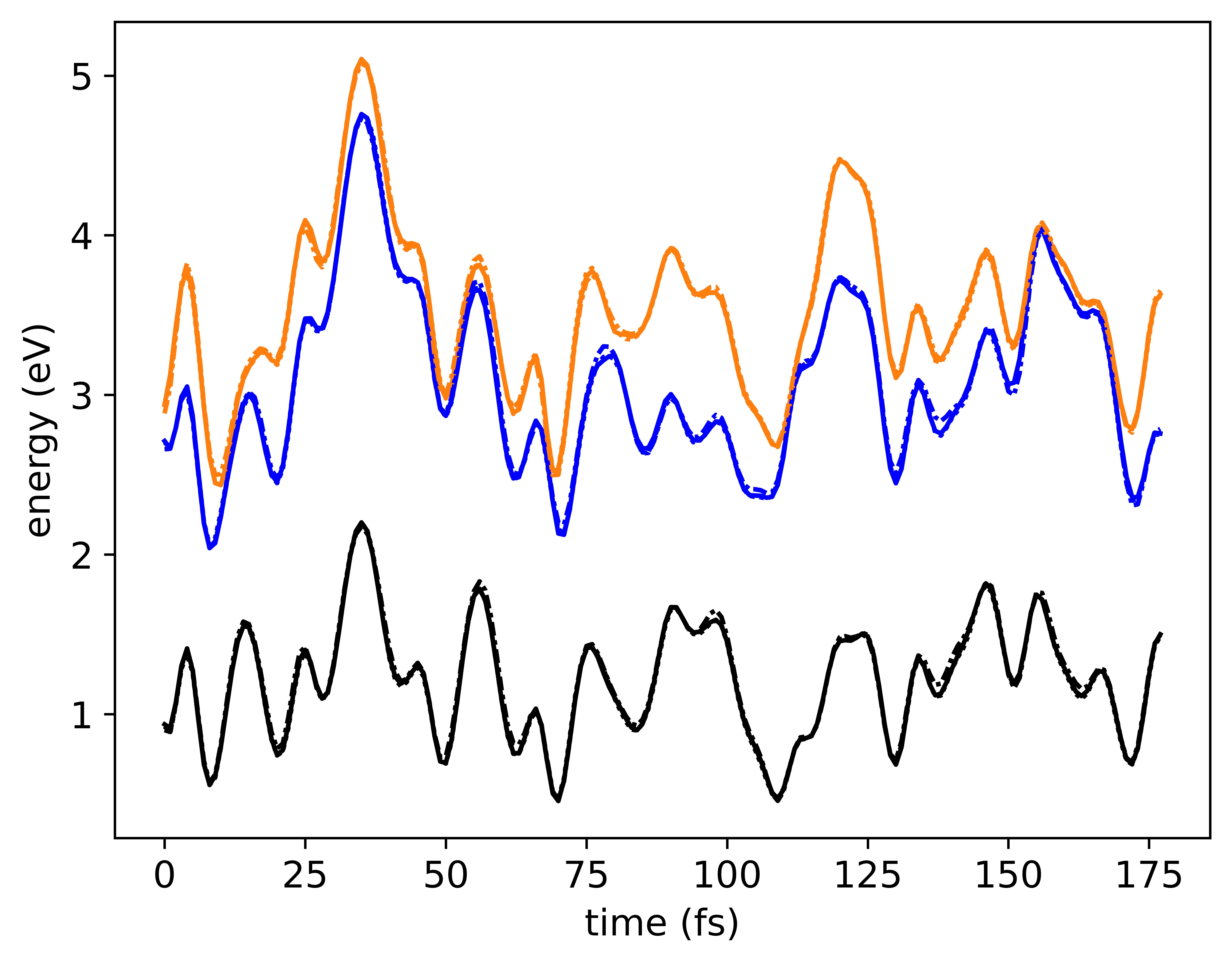}}%
\subfigure{\label{fig:cc}\includegraphics[width=0.35\textwidth]{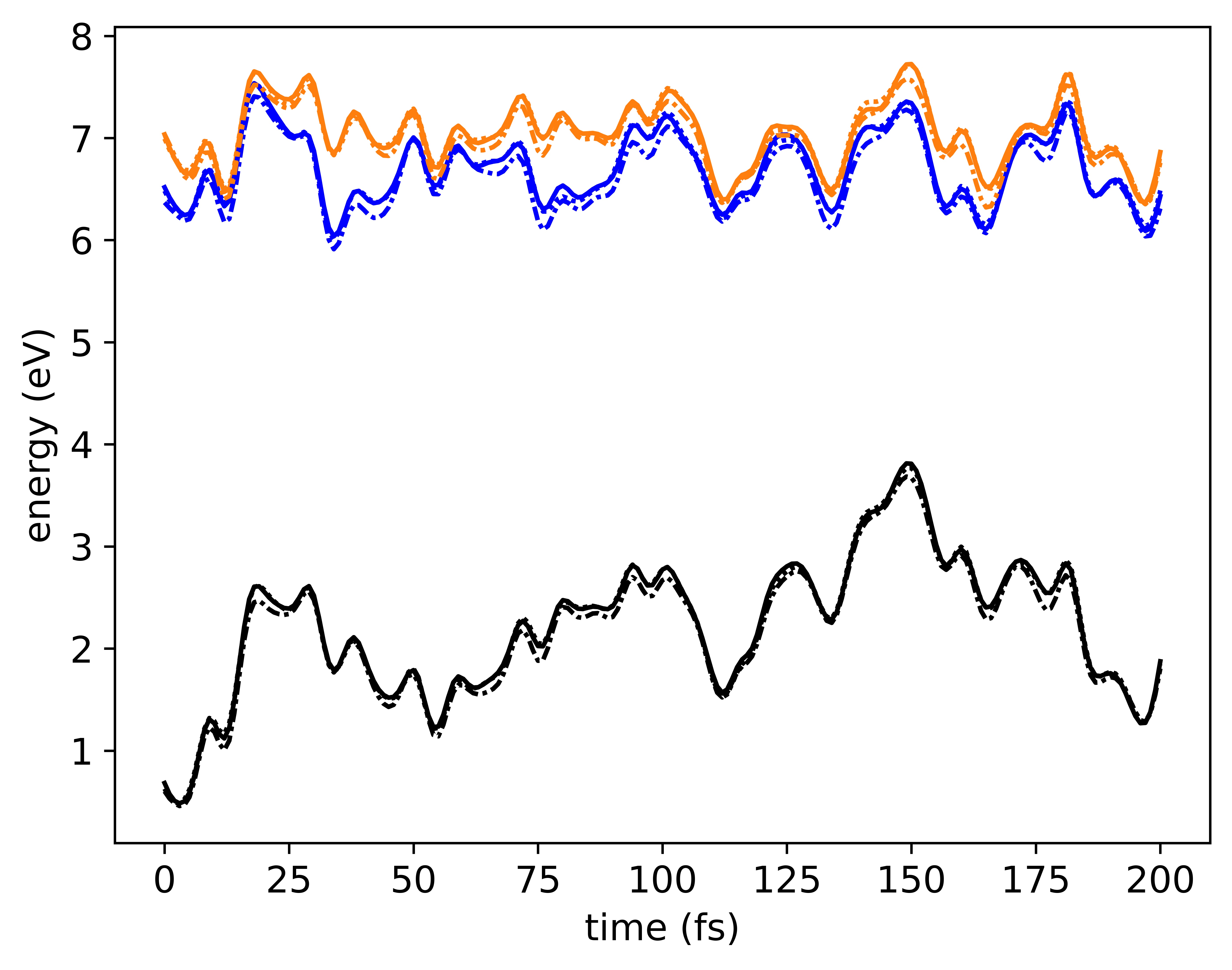}}%
}
\caption{Comparison of neural network generated PESs with full data(dashed), and halved data (dotted) for the best network found by the GS algorithm for Pyrazine (left), for Pyrrole (middle), and for Furan (right) and reference ab initio based PESs (solid). $S_0$ (black); $S_1$ (blue); $S_2$ (orange).}

\label{fig:trajectories}
\end{figure}

\begin{figure}[H]
\makebox[\linewidth][c]{%
\centering
\subfigure{\label{fig:aaa}\includegraphics[width=0.35\textwidth]{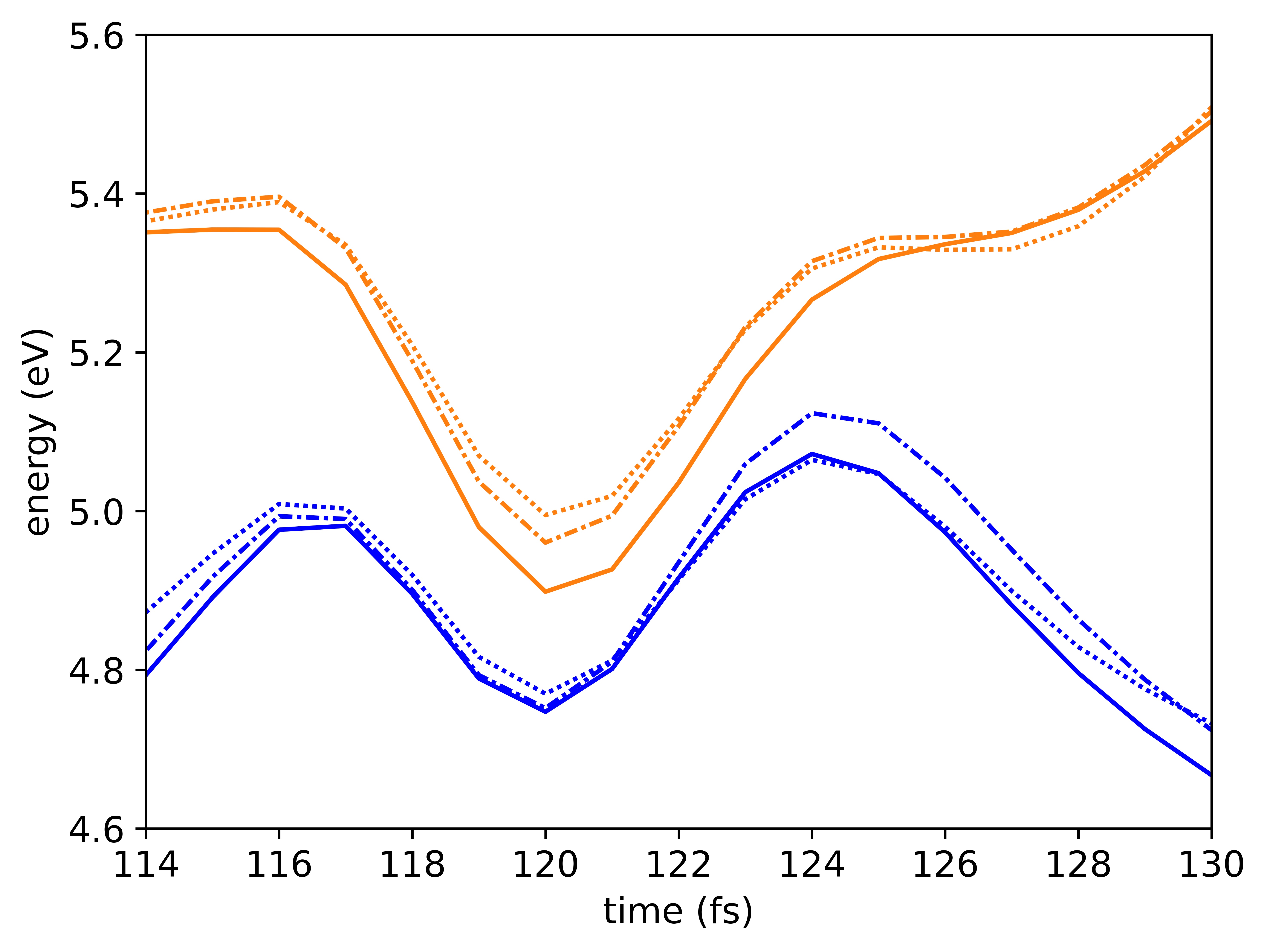}}
\subfigure{\label{fig:bbb}\includegraphics[width=0.35\textwidth]{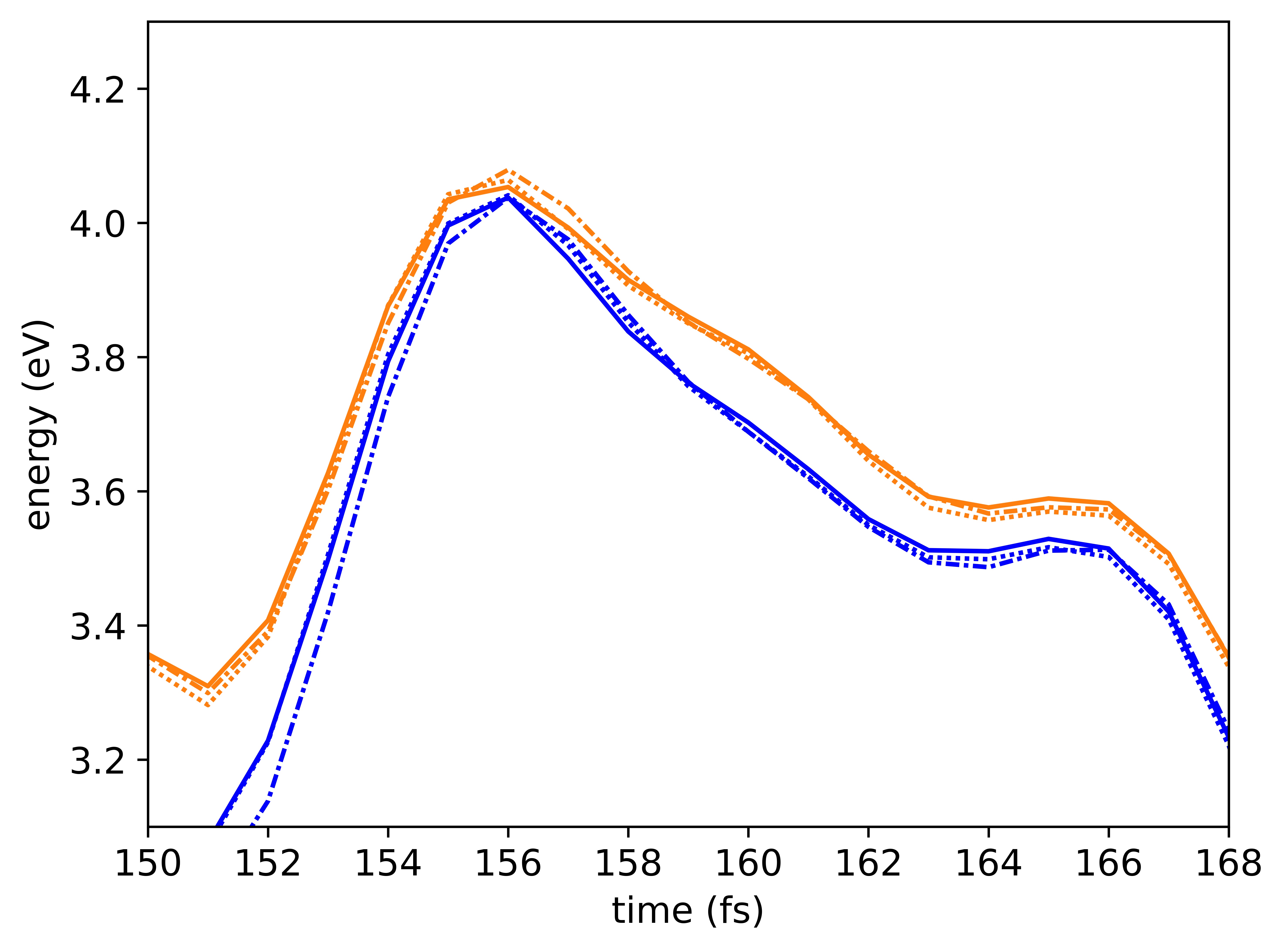}}%
\subfigure{\label{fig:ccc}\includegraphics[width=0.35\textwidth]{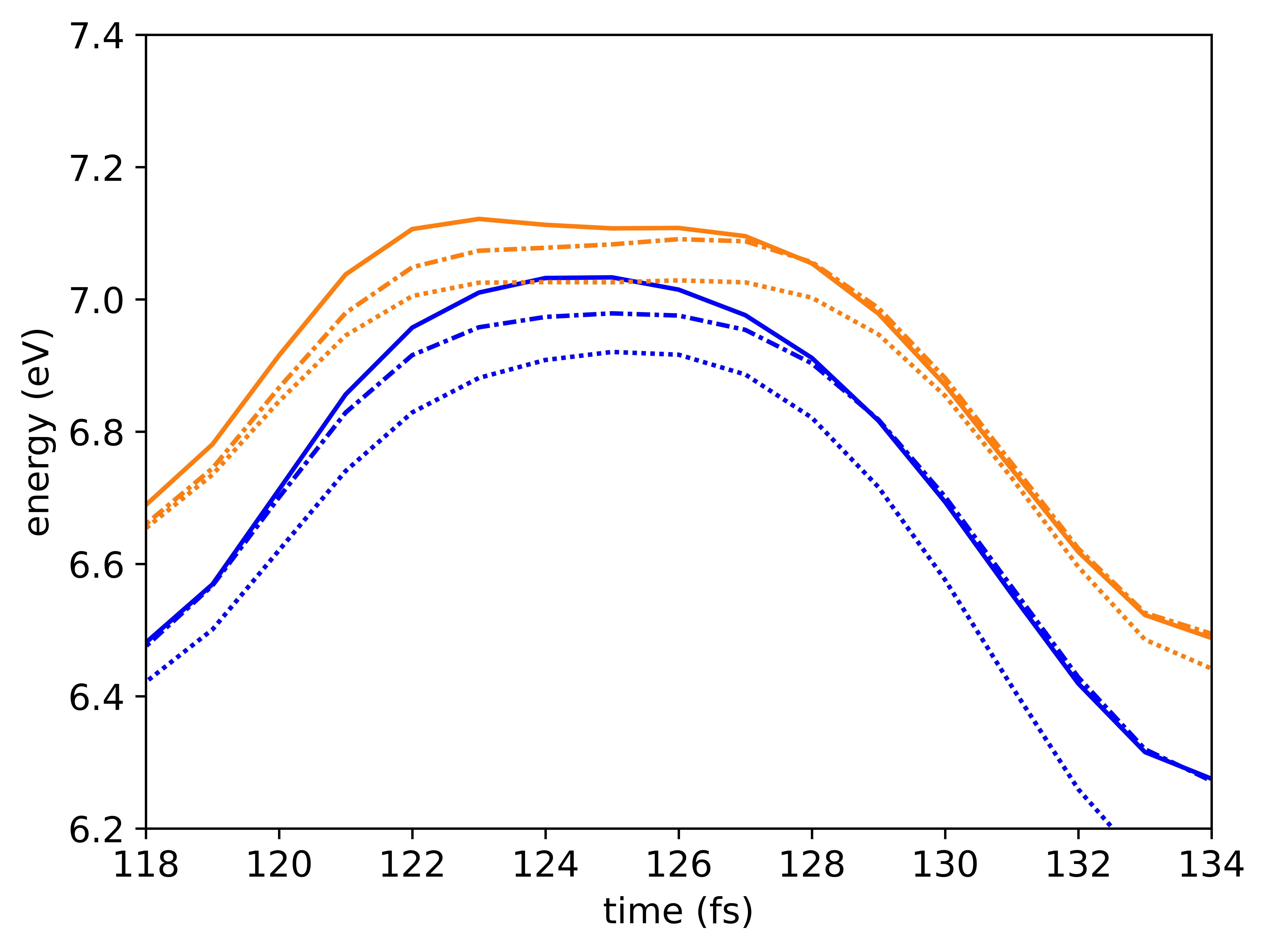}}%
}
\caption{Comparison of neural network generated PESs with full data(dashed), and halved data (dotted), for the best network found by the GS algorithm for Pyrazine (left), for Pyrrole (middle), and Furan (right) and and reference ab initio based PESs (solid), zoomed in on the areas where the $S_1$ and $S_2$ surfaces are close. $S_1$ (blue); $S_2$ (orange).}
\label{fig:trajectorieszoom}
\end{figure}

\subsection{Training error over time}

\begin{figure}[H]
\makebox[\linewidth][c]{%
\centering
\subfigure{\label{fig:aa}\includegraphics[width=0.35\textwidth]{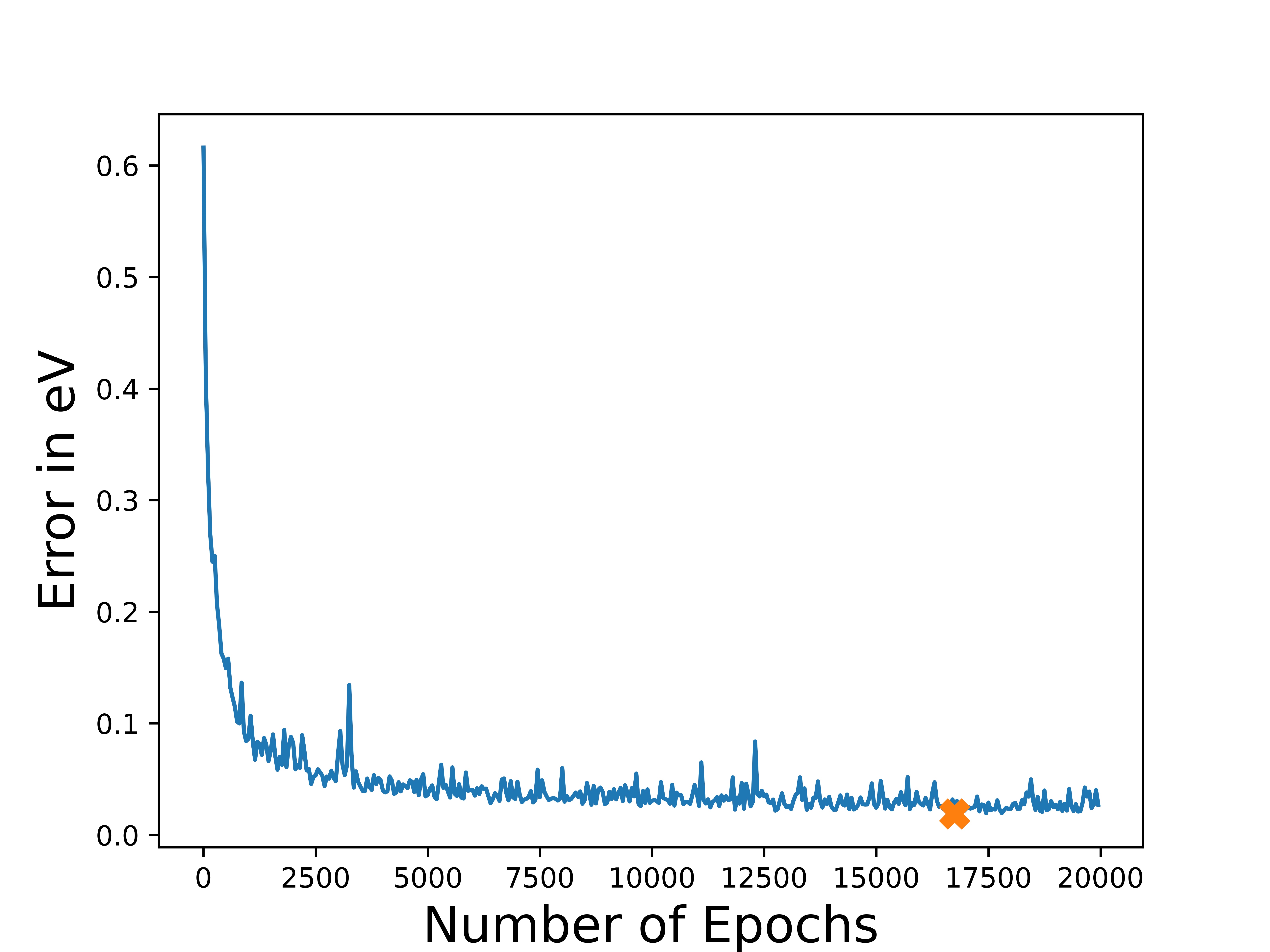}}%
\subfigure{\label{fig:bb}\includegraphics[width=0.35\textwidth]{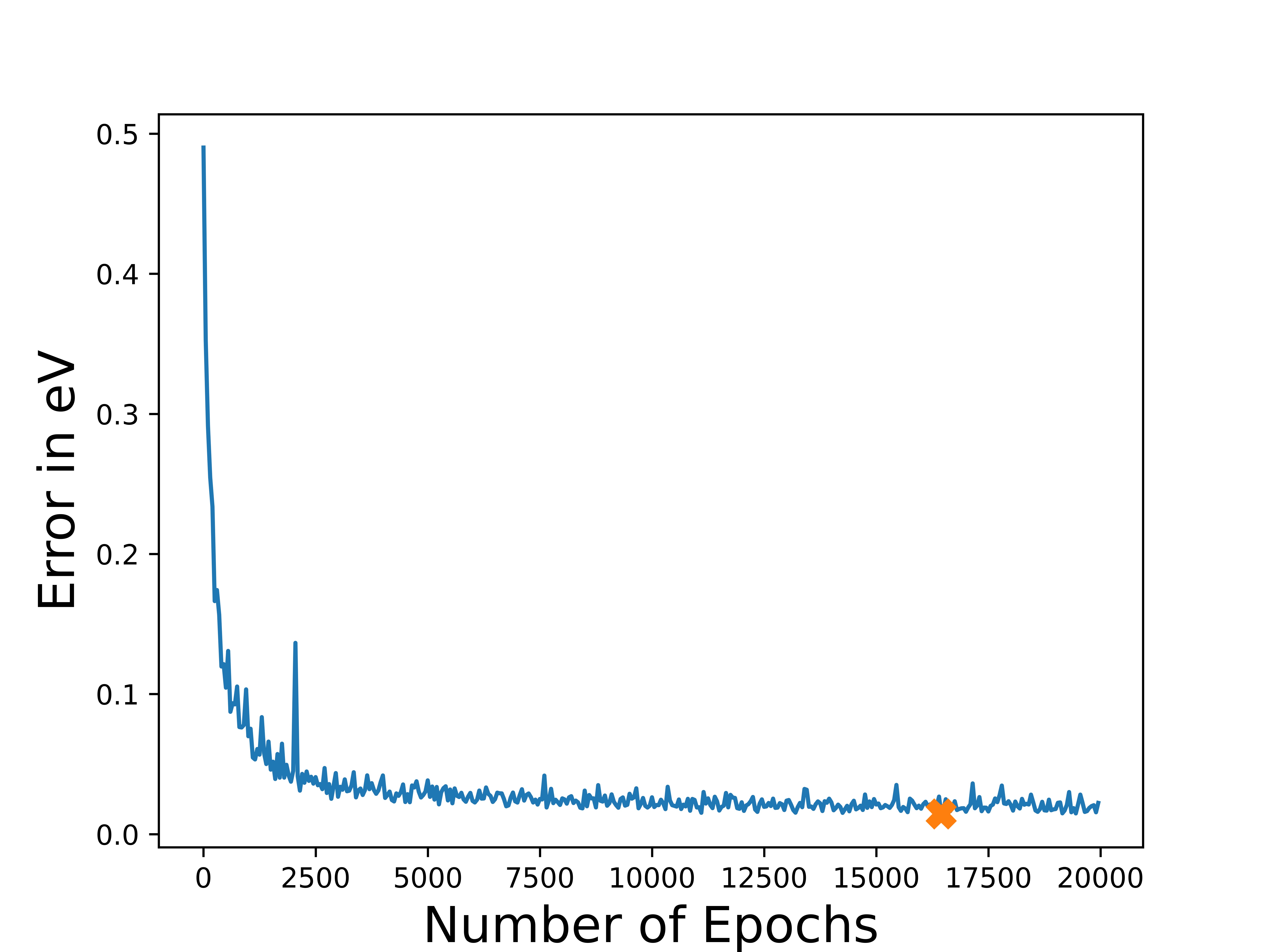}}%
\subfigure{\label{fig:cc}\includegraphics[width=0.35\textwidth]{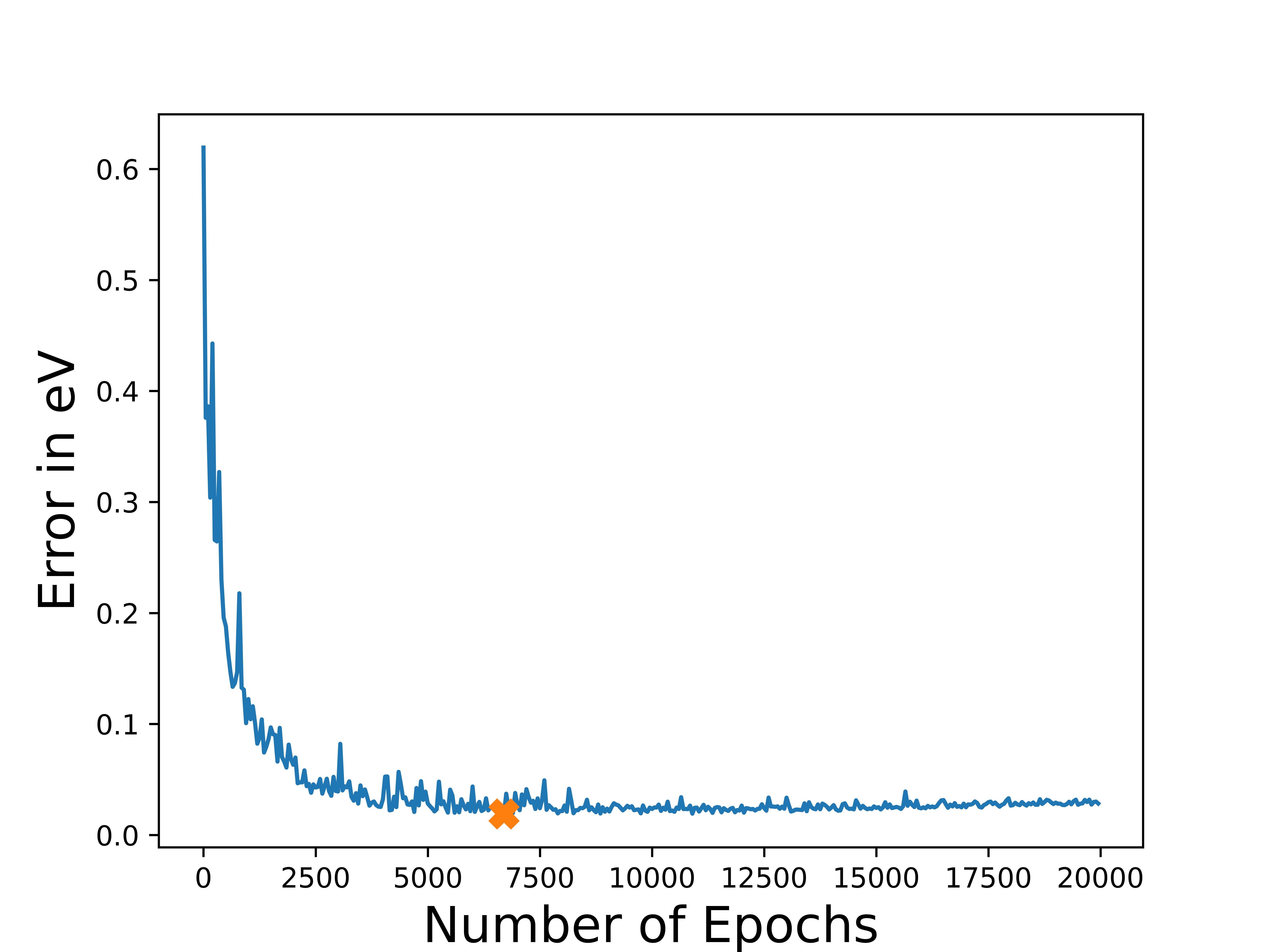}}%
}
\caption{Validation error during a training run using the optimal hyperparameters found by GS for Pyrazine, Pyrrole and Furan. The orange cross represents the lowest validation error found during the training process.}
\label{fig:validationrun}
\end{figure}

To give an account of a single training run in HOAX, the trajectories of the validation error during a training run of 20.000 is given in Figure \ref{fig:validationrun}. The decrease in error is small after the cutoff point of 10.000 epochs used in the hyperparameter optimization

